\documentclass[11pt]{article}

\pdfoutput=1 
\usepackage{jheppub}

\usepackage{amsmath,amssymb,amsfonts}
\usepackage{hyperref}
\hypersetup{colorlinks,citecolor=blue,urlcolor=blue,linkcolor=blue}
\usepackage{graphicx}
\usepackage{xcolor}

\usepackage{caption}
\usepackage{subcaption}

\usepackage{nicefrac}
\usepackage{mathrsfs}
\usepackage{mdframed}
\usepackage{ulem}
\usepackage{dsfont}
\usepackage{units}
\usepackage[english]{babel}
\usepackage{float} 
\usepackage{braket}
\def\beq{\begin{equation}\begin{aligned}}
\def\eeq{\end{aligned}\end{equation}}

\begin{document}

\title{How to Count States in Gravity}

\author[1,2,3]{Vijay Balasubramanian}
\author[4]{Tom Yildirim}
\affiliation[1]{Department of Physics and Astronomy, University of Pennsylvania, Philadelphia, PA 19104, U.S.A.}
\affiliation[2]{Theoretische Natuurkunde, Vrije Universiteit Brussel and International Solvay Institutes, Pleinlaan 2, B-1050 Brussels, Belgium}
\affiliation[3]{Rudolf Peierls Centre for Theoretical Physics, University of Oxford,Oxford OX1 3PU, U.K.}
\affiliation[4]{Department of Physics, Keble Road, University of Oxford, Oxford, OX1 3RH, UK}
\emailAdd{vijay@physics.upenn.edu}
\emailAdd{tom.yildirim@physics.ox.ac.uk}

\abstract{
Gibbons and Hawking proposed that the Euclidean gravity path integral with periodic boundary conditions in time computes the thermal partition sum of gravity.  As a corollary, they argued that a derivative of the associated free energy with respect to the Euclidean time period computes gravitational entropy.  Why is this interpretation correct?  That is, why does this path integral compute a trace over the Hilbert space?   Here, we show that the quantity computed by the Gibbons-Hawking path integral is equal to an {\it a priori} different object -- an explicit thermal trace over the  Hilbert space spanned by states produced by the Euclidean gravity path integral.   This follows in two ways: (a) if the Hilbert space  with two boundaries factorizes into a product of two single boundary Hilbert spaces, as we have previously shown; and (b) via explicit resolution of the trace by a spanning basis of states.  We similarly show how a replicated Euclidean  gravity path integral with a single periodic boundary computes a Hilbert space trace of powers of the density matrix, explaining why this approach computes the entropy of states entangled between two universes.
}

\maketitle
\newpage

\section{Introduction}

The Bekenstein-Hawking  entropy formula
$ \mathbf{S}_{BH} = \frac{A}{4\mathrm{G_N}}$ is believed to universally relate the number of microstates of a  black hole to its horizon area $A$. This formula was  proposed  through an analogy between classical laws of black hole mechanics and the laws of thermodynamics \cite{Bekenstein:1973ur,Bardeen:1973gs}. The analogy was made literal by Hawking's result \cite{Hawking:1975vcx} that black holes evaporate by emitting (seemingly) thermal radiation at temperature $T_H = \frac{1}{8\pi M \mathrm{G_N}}$, derived by considering quantum field theory on black hole backgrounds. Thus it appears that black holes are finite dimensional quantum systems with Hilbert spaces of dimension $e^{\mathbf{S}_{BH}}$. 

A second argument to the same effect was proposed by Gibbons and Hawking \cite{Gibbons:1976ue}. They suggested that the Euclidean gravitational path integral with periodic boundary conditions in time should be interpreted as the thermal partition sum $Z$ of gravity.  This is in analogy with statistical field theory where the partition sum is, by construction, computed by the Euclidean path integral with a periodic time $\beta$ equal to the inverse temperature.   Leaving aside some technical subtleties, the leading saddlepoint of the Gibbons-Hawking path integral turns out to be the Euclidean black hole, and so in the semiclassical limit $Z \approx e^{-I_{BH}}$ where $I_{BH}$ is the black hole action.  Pursuing the analogy with thermodynamics, we infer that the entropy is $S = (1-\beta \partial_\beta )\ln{Z}$, where $\beta$ is the Euclidean time period.  This computation reproduces the Bekenstein-Hawking entropy formula.

But does the Gibbons-Hawking path integral actually compute a thermal trace over the Hilbert space?  In  statistical physics the path integral computing this trace is performed over a cylinder, i.e., a manifold of the form ${\cal M} \times S^1$ where the  period of the circle is the inverse temperature $\beta$.  But in gravity we must  sum over all topologies and geometries that are asymptotic to $\partial {\cal M} \times S^1$.  In fact, in the Euclidean black hole saddlepoint which dominates the path integral in appropriate regimes, the $S^1$ is contractible in the interior -- the topology is that of a disc, not a cylinder.   To take the $\partial/\partial \beta$ derivative required for computing the entropy we have to vary the period, and if we do this while keeping the rest of the saddlepoint metric fixed, a conical singularity develops at the origin.  This off-shell geometry then has the topology of a cylinder, but it is not clear why this is sufficient to admit a trace interpretation of the path integral.  More generally, it is not clear  whether the Gibbons-Hawking path integral computes a trace over some Hilbert space,  what this Hilbert space might be, and whether this Hilbert space  spans the possible states of quantum gravity.

Here, we establish that the Gibbons-Hawking path integral equals the thermal trace $Tr_{\mathcal{H}_{\mathcal{X}}}(e^{-\beta H})$
over the Hilbert space $\mathcal{H}_{\mathcal{X}}$ of quantum gravity with a single boundary. We show this in two ways by using the methods of \cite{Balasubramanian:2025jeu}.  First, we show that the result follows if the Hilbert space of quantum gravity with two boundaries factorizes into a product of Hilbert spaces with one boundary, as we have recently demonstrated \cite{Balasubramanian:2025zey}. In detail, we strategically insert a novel resolution of the identity in the two-boundary gravity Hilbert space into the Euclidean path integral with periodic time  \cite{Balasubramanian:2025jeu}. The terms in the resulting sum are  equal to the terms in the thermal trace over the single boundary Hilbert space. Second, we use a tractable basis of states for the single boundary Hilbert space  \cite{Balasubramanian:2025zey} to explicitly evaluate the path integrals in the thermal trace, and show that their sum collapses to a Euclidean path integral with a single periodic boundary.\footnote{These single-sided shell states are a single-boundary version of the two-sided shell states used  to provide a microscopic understanding of black hole entropy in \cite{Sasieta:2022ksu,Balasubramanian:2022gmo,Balasubramanian:2022lnw,Climent:2024trz,Balasubramanian:2024rek}. They were described in \cite{Keranen:2015fqa,Chandra:2022fwi}.}   In the regime of parameters  where the leading saddlepoint of the Gibbons-Hawking path integral is a Euclidean black hole, these calculations give a  micro-canonical Hilbert space dimension of $e^{A/4GN}$ where $A$ is the horizon area  after Lorentzian continuation of the saddlepoint.  However, as we will discuss, the basis of states explaining this dimension need not have horizons at all.

We then consider states of universes with two asymptotic boundaries.  Since the Hilbert space of such universes is a product of factors  \cite{Balasubramanian:2025zey}, we can determine the reduced density matrix on either factor by tracing out the other.   The R\'{e}nyi entropies should then be computed  as a trace of powers of this density matrix, and the von Neumann entropy is recovered as a limit.  However, the conventional approach instead computes a replicated Euclidean gravity path integral with a single periodic boundary, by analogy with the Gibbons-Hawking proposal for the gravitational partition sum.   We show why this replicated integral is equal to the sum over path integrals that calculates the Hilbert space trace required for the R\'{e}nyi entropy.

Four sections follow. In Sec.~\ref{sec:QTFTPI} we review how  bases of states can be constructed by cutting open the gravitational path integral. In Sec.~\ref{sec:TPF} we show why the Gibbons-Hawking path integral computes a thermal trace over the Hilbert space. In Sec.~\ref{sec:ER=EPR} we explain why a replica path integral can evaluate the trace required to compute gravitational entanglement entropy. Finally, we conclude with a  discussion in Sec.~\ref{sec:Summary}.

\section{States from the gravitational path integral} \label{sec:QTFTPI}
We would like to compute the thermal partition function of a gravity theory 
 obtained by tracing over an orthonormal basis of states $Tr_{\mathcal{H}_{\mathcal{X}}}(e^{-\beta H})= \sum_{\gamma} \langle v_\gamma| e^{-\beta H}|v_\gamma\rangle$.  We need three ingredients: ({\bf 1}) A spanning  set of states which could be overcomplete; ({\bf 2}) The action of the Hamiltonian on these states; ({\bf 3}) A way of taking the trace over the Hilbert space. As we will explain, all three are naturally available for states defined by cutting open the Euclidean gravitational path integral.

\begin{figure}[h]
    \centering
    \includegraphics[width=0.5\linewidth]{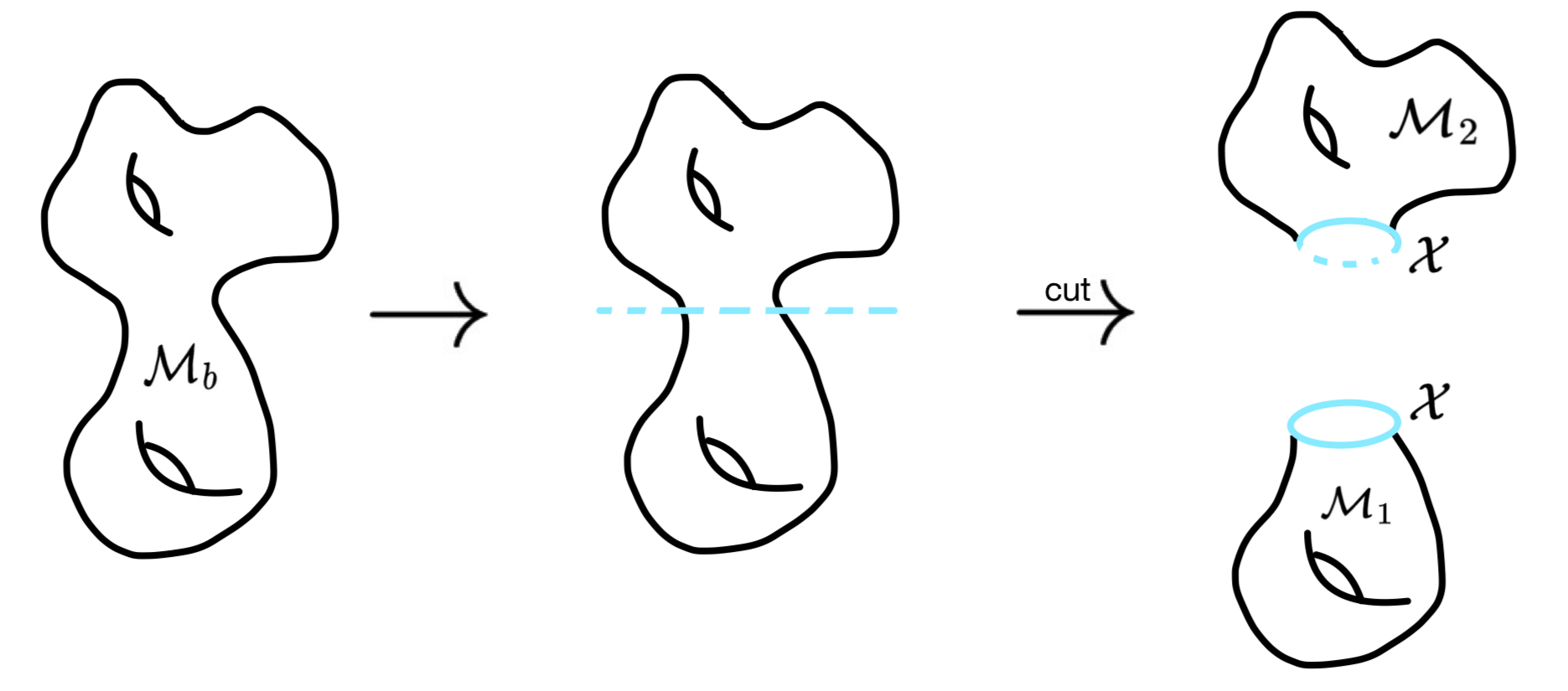}
    \caption{Slicing a gravity path integral to define a state (adapted from \cite{Balasubramanian:2025jeu}).}
    \label{fig:cutpathint}
\end{figure}

\paragraph{How to make states.}First we review the construction of states by cutting open the Euclidean gravitational path integral, understood as an effective description below a cutoff, completed at high energy by some fundamental theory. See \cite{Balasubramanian:2025jeu, Balasubramanian:2025zey} for details.  Our considerations will apply to universes with either vanishing or negative cosmological constant.  The path integral computes a sum  over  topologies and geometries ($g$), and matter fields ($\phi$), compatible with specified asymptotic boundary conditions (boundary topology, metric, and matter sources, collectively denoted by $\Sigma_b$): $\zeta[\Sigma_b] = \int_{g,\phi \to \Sigma_b} \mathcal{D}g\mathcal \, \mathcal{D}\phi \, e^{-I_{grav}[g,\phi]}$, where $I_{grav}$ is the gravitational and matter action.  To  construct states we partition the  boundary  $\mathcal{M}_b =\mathcal{M}_1 \cup \mathcal{M}_2 $ so that $\partial\mathcal{M}_1= \partial\mathcal{M}_2=\mathcal{X}$ (Fig.~\ref{fig:cutpathint}). The resulting functional of the fields and geometry on the cut defines states  $|\mathcal{M}_{1,2}\rangle \in \mathcal{H}_{\mathcal{X}}$ in a Hilbert space $\mathcal{H}_{\mathcal{X}}$ such that $\langle \mathcal{M}_{2}|\mathcal{M}_{1}\rangle=\zeta[\Sigma_b]$. We generally resort to saddlepoint approximations to evaluate  overlaps of these states.
Given a cut $\mathcal{X}$ of a boundary manifold $\mathcal{M}_b$ we can produce different states $\ket{i}$ by inserting matter operators $\mathcal{O}_i$ on $\mathcal{M}_{b}$. To compute overlaps $\braket{i|j}$ we evaluate the path integral with $\mathcal{O}_{i,j}$ inserted on $\mathcal{M}_{1,2}$.  The bra  $\bra{i}$ associated to the ket $\ket{i}$ defined by inserting  $\mathcal{O}_i$  on $\mathcal{M}_{1}$ is obtained by inserting  $\mathcal{O}^{\dagger}_i$ on $\mathcal{M}_{1}$. The ket $\ket{i^*}$  defined inserting $\mathcal{O}^{\dagger}_i$ on $\mathcal{M}_{1}$ therefore defines the same boundary condition as $\bra{i}$, and so  $\langle i|j\rangle = \langle j^*|i^*\rangle$.  Finally, the leading saddlepoint of the path integral for the norm of a state can be sliced open at a moment time-reflection symmetry, provided this exists, and analytically continued to Lorentzian signature.  This procedure associates a Lorentzian geometry to the state. 

\paragraph{Factorisation.} We want to construct states in the Hilbert space of quantum gravity with two disconnected Lorentzian boundaries.  To do so we slice open a Euclidean path integral so that cut on the boundary  has two disconnected components  $\mathcal{X}_L \cup \mathcal{X}_R $.  As discussed in \cite{Balasubramanian:2025zey}, there are two  ways of  constructing  states in this Hilbert space.  First, we could slice open the Euclidean path integral with two separate closed boundaries,  one with cut $\mathcal{X}_L $ and the other with $\mathcal{X}_R$. This produces states in a tensor product Hilbert space,  $\mathcal{H}_{\mathcal{X}_L}\otimes \mathcal{H}_{\mathcal{X}_R}$. Alternatively, we can slice open the path integral with a single  closed connected boundary manifold with cut  $\mathcal{X}_L \cup \mathcal{X}_R $.  This procedure produces intrinsically two sided states making up a Hilbert space  $\mathcal{H}_{LR}$.  Naively,  $\mathcal{H}_{\mathcal{X}_L \cup \mathcal{X}_R} = \mathcal{H}_{LR} \,  \cup \, \mathcal{H}_{\mathcal{X}_L}\otimes \mathcal{H}_{\mathcal{X}_R}$.  However, the authors of \cite{Balasubramanian:2025zey} computed overlaps between states in  $ \mathcal{H}_{\mathcal{X}_L \cup\mathcal{X}_R}$ and tensor product states in $\mathcal{H}_{\mathcal{X}_L}\otimes \mathcal{H}_{\mathcal{X}_R}$ by gluing the corresponding path integrals together along the cuts.\footnote{In two-dimensional JT gravity, \cite{Boruch:2024kvv} showed that traces in the two-sided Hilbert space factorize into a product of two contributions to all orders in $e^{-1/G_N}$. Subsequently, \cite{Balasubramanian:2024yxk} extended this result to arbitrary spacetime dimensions within the leading saddle-point approximation. A key ingredient in the arguments presented here is the identification of these two factors with the single-boundary gravitational Hilbert space, a correspondence established in \cite{Balasubramanian:2025zey}.} They showed that $\mathcal{H}_{LR} = \mathcal{H}_{\mathcal{X}_L}\otimes \mathcal{H}_{\mathcal{X}_R}$, so that the complete two-boundary Hilbert space factorises: 
\beq \label{eq:facprob}
\mathcal{H}_{\mathcal{X}_L \cup\mathcal{X}_R}=\mathcal{H}_{\mathcal{X}_L}\otimes \mathcal{H}_{\mathcal{X}_R} \, .
\eeq

\begin{figure}[h]
    \begin{subfigure}[b]{0.2\linewidth}
        \centering
        \includegraphics[width=\linewidth]{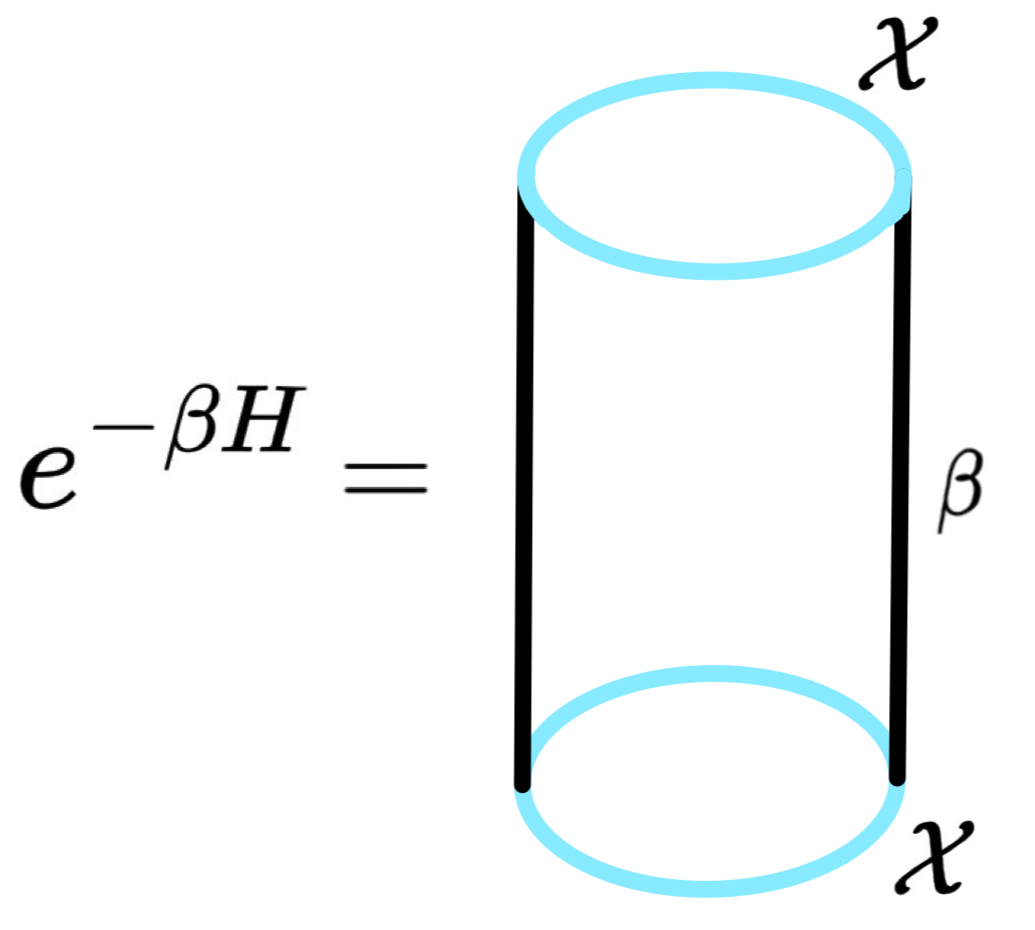}
      \caption{}
        \label{fig:hamiltonian}
    \end{subfigure}
    \begin{subfigure}[b]{0.2\linewidth}
        \centering
        \includegraphics[width=\linewidth]{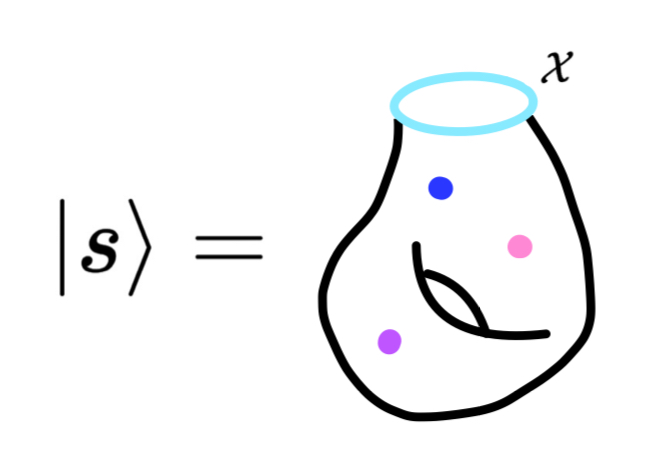}
        \caption{}
        \label{fig:state}
    \end{subfigure}
    \hfill
    \begin{subfigure}[b]{0.24\linewidth}
        \centering
        \includegraphics[width=\linewidth]{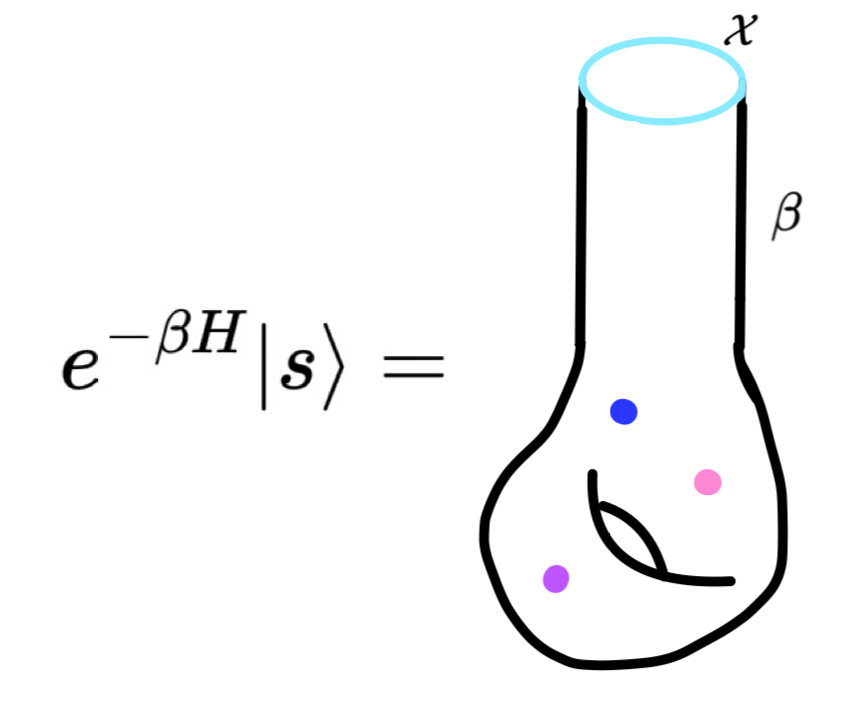}
        \caption{}
        \label{fig:timeEstate}
     \end{subfigure}
     \hfill
     \begin{subfigure}[b]{0.24\linewidth}
        \centering
        \includegraphics[width=\linewidth]{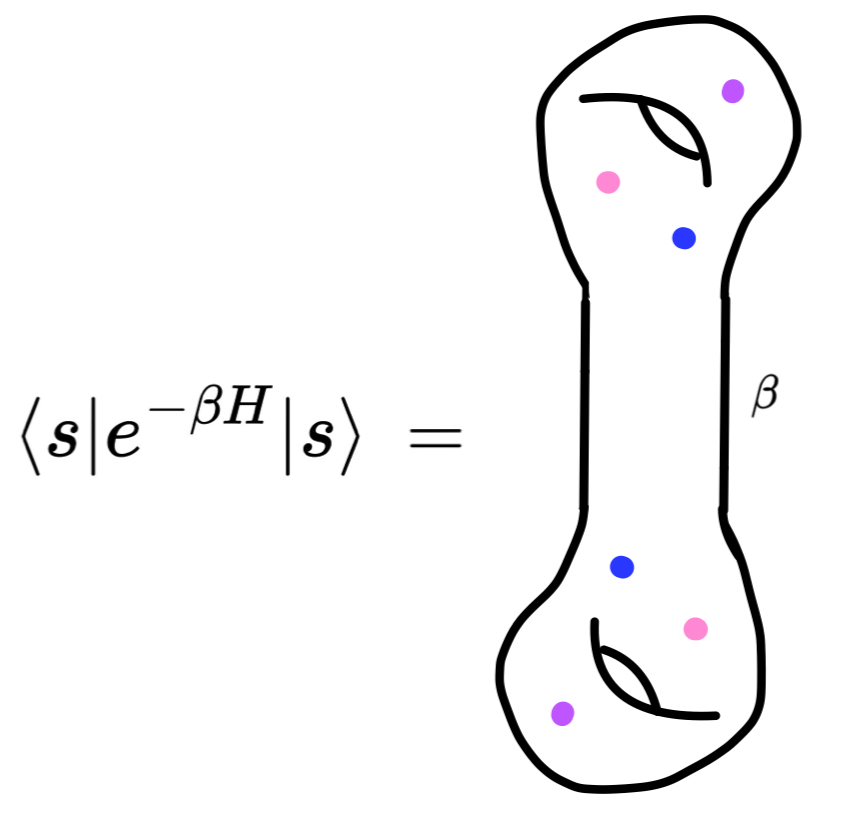}
        \caption{}
        \label{fig:timeEoverlap}
     \end{subfigure}
    \caption{Action of $e^{-\beta H}$ on states.  ({\bf a}) Action of the Hamiltonian, from a different perspective this is preparing the state $|2\beta\rangle$. ({\bf b}) Example state in $\mathcal{H}_{\mathcal{X}}$. ({\bf c}) Action of the Hamiltionian on this state. ({\bf d}) Survival amplitude after Euclidean time evolution, alternatively the overlap $\langle{2\beta}\ket{s^*}_L\otimes\ket{s}_R$. (adapted from \cite{Balasubramanian:2025jeu}). }
\end{figure}

\paragraph{Action of the Hamiltonian.} \label{sec:BHentropy}
Let $H$ be the Hamiltonian.  Then $e^{-\beta H}$ acts to glue an asymptotic boundary ``cylinder"  $\mathcal{X} \times \mathbb{I_{\beta}}$ to the cut (Fig.~\ref{fig:hamiltonian},\ref{fig:state},\ref{fig:timeEstate}).
So the amplitude  $\langle S| e^{-\beta H}|S\rangle$  is constructed by sewing the $ |S \rangle$ bra and ket together via a boundary cylinder of length $\beta$ (Fig.~\ref{fig:timeEoverlap}).  The action of $e^{-\beta H}$ can also be viewed as creating a state  defined on a cut consisting of two copies of $\mathcal{X}$ (think of folding the cylinder around to define a path integral sliced open to obtain a state in the two-boundary Hilbert space $\mathcal{H}_{\mathcal{X}_L \cup\mathcal{X}_R}$.  We will say that   $|\beta\rangle$ is a state produced in this way by  the action of $e^{-\frac{\beta H}{2}}$, so that the norm $\braket{\beta|\beta}$ is given by a path integral with boundary $\mathcal{X} \times S^1$ where the $S^1$ has period $\beta$. From this point of view, we can also think of overlaps between time evolved states in $\mathcal{H}_{\mathcal{X}}$, for example $\bra{i}^{-\frac{\beta H}{2}}\ket{j}$, as overlaps between  $|\beta\rangle$ and a tensor product state, $\langle{\beta}\ket{i^*}_L\otimes\ket{j}_R$.

\paragraph{Wormholes and coarse-graining.} To compute the squared overlap, $\braket{i|j}\braket{j|i}$ we insert $\mathcal{O}^{(\dagger)}_{i,j}$ on two copies of the boundary manifold and sum over  bulk geometries and topologies that fill in these  boundaries. This path integral has  two kinds of contributions: (a) disconnected topologies coming from products of terms contributing to  $\braket{i|j}$, and (b) connected topologies, i.e., wormholes, that interpolate between the two boundaries. The latter lead an apparent paradox: $\braket{i|j}= \delta_{ij}$ while $\braket{i|j}\braket{j|i}= \delta^2_{ij} + Z_{WH}$, where $Z_{WH}$ is the wormhole contribution.  Following
\cite{Saad:2019lba,Saad:2019pqd,Sasieta:2022ksu,Marolf:2020xie,deBoer:2023vsm,deBoer:2024mqg,Penington:2019kki,Almheiri:2019qdq,Balasubramanian:2022gmo,Balasubramanian:2022lnw,Climent:2024trz,Balasubramanian:2024yxk,Marolf:2024jze,Balasubramanian:2025zey,Balasubramanian:2025jeu}
we interpret this as saying that the gravitational path integral computes coarse-grained averages over  quantities in an underlying fine-grained theory. We therefore write an overbar for quantities computed from the path integral, e.g., $\overline{\braket{i_1|j_1}\braket{i_2|j_2} \cdots \braket{i_n|j_n}}$ denotes the  path integral evaluation of the product of $n$ overlaps.

\paragraph{Taking traces in an overcomplete basis.} 
To compute the thermal partition function we must perform a trace, for which it is helpful to have a basis of states. Generic bases made by slicing the gravity path integral will not be orthogonal because of  the  wormhole contributions to the square of the overlap described above. Also it is often convenient in gravity to construct over-complete, and therefore non-orthogonal, bases \cite{Penington:2019kki,Balasubramanian:2025zey,Balasubramanian:2025jeu}.  Let $\{ \ket{i} \}$ be such a basis with a Gram matrix of overlaps $G_{ij} \equiv \braket{i|j}$.  Then we can construct an orthonormal basis  $|v_\gamma\rangle= G^{-1/2}_{ji}U_{i\gamma}|j\rangle$; here $U$ is a unitary transformation  diagonalizing the Gram matrix ($G=UDU^{\dagger}$ with $D$ diagonal), where from now on repeated indices are understood as summed (details in Appendix~A of \cite{Balasubramanian:2024yxk,Balasubramanian:2025jeu} ).  
Here $G^{-1}$ is defined by analytic continuation of $(G^{n})_{ij}$ from positive integer $n$: $ G^{-1}_{ij} \equiv \lim_{n\to -1} (G^{n})_{ij}$. This expression inverts  nonzero eigenvalues of $G$ and leaves  zero eigenvalues untouched. We can now write the trace as
\beq 
\label{eq:tr}Tr_{\mathcal{H}_{\mathcal{X}}}(O) = G^{-1}_{ij}\langle j|O|i\rangle \, \equiv  \lim_{n\to -1} (G^{n})_{ij}\langle j|O|i\rangle \, ,
\eeq
and resolve the identity on $\mathcal{H}_{\mathcal{X}}$ as
\beq\label{eq:id}
\mathds{1}_{{\mathcal{X}}} = \sum_{\gamma}|v_\gamma\rangle\langle v_\gamma| = G^{-1}_{ij} |i\rangle\langle j| \equiv \lim_{n\to -1} (G^{n})_{ij}|i\rangle\langle j| \, .
\eeq

\paragraph{A toolkit for fine-grained equalities.}
Suppose we want to use the gravitational path integral to show that $\mathcal{A} = \mathcal{B}$ in the fine-grained theory.  Since the path integral only computes coarse grained averages $\bar{\mathcal{A}}$ and $\bar{\mathcal{B}}$ as discussed above, we have to show both $\bar{\mathcal{A}} - \bar{\mathcal{B}} = 0$ and 
 $\overline{(\mathcal{A}-\mathcal{B})^2} = 0$ as the latter means that $\mathcal{A} - \mathcal{B} = 0$ for all the fine-grained configurations that are being averaged.    In general we can only evaluate the path integral computing these quantities explicitly in the saddlepoint approximation, but, as pointed out in \cite{Balasubramanian:2025jeu}, if we  want to show  $\overline{\mathcal{A}} = \overline{\mathcal{B}} $  it suffices to show that each contribution to the path integral $\overline{\mathcal{A}}$ is matched by an equal contribution to $\overline{\mathcal{B}}$ and vice versa.  We can achieve this in three steps \cite{Balasubramanian:2025jeu}: ({\bf 1}) Construct an overcomplete basis of states by inserting operators into the Euclidean path integral;  ({\bf 2}) Take a limit in which the basis is infinitely overcomplete, leading to drastic simplifications of the topologies contributing to amplitudes;   ({\bf 3}) Demonstrate an action-preserving bijection between geometries contributing to the path integrals of interest after this topological simplification. This task can be simplified by insertions of the identity (\ref{eq:id}).  The authors of \cite{Balasubramanian:2025jeu} used this procedure to demonstrate that the ``two-sided shell states'' of 
\cite{Sasieta:2022ksu,Balasubramanian:2022gmo,Balasubramanian:2022lnw,Antonini:2023hdh,Balasubramanian:2024rek}  form an over-complete basis for quantum gravity with two boundaries. These methods can also be used to demonstrate a basis for quantum gravity with one boundary, and that the two-boundary Hilbert space factorizes into a product of two one-boundary Hilbert spaces \cite{Balasubramanian:2025zey}, a fact that we will use.

\subsection{The Gibbons-Hawking puzzle}\label{sec:puzzles}

\begin{figure}[h]
    \begin{subfigure}{\linewidth}
    \centering
    \includegraphics[width=0.7\linewidth]{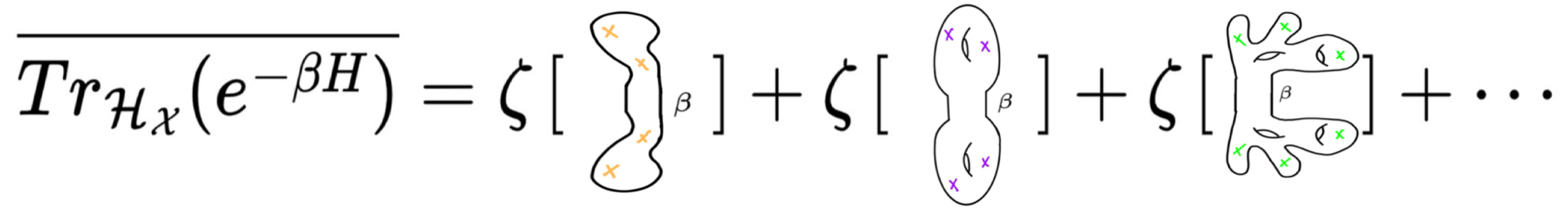}
    \caption{
    }
    \label{fig:gravpartition}
     \end{subfigure}
 \begin{subfigure}{\linewidth}
      \centering
    \includegraphics[width=0.25\linewidth]{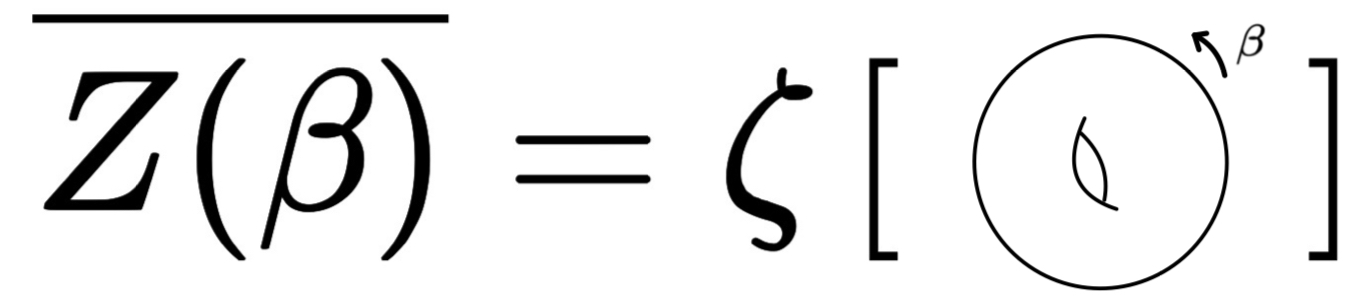}
    \caption{
    }
    \label{fig:BHassump}
 \end{subfigure}
     \caption{({\bf a})  The thermal partition function in gravity computed by summing over  path integrals with boundaries defined by the basis of states. The colored crosses denote  possible matter insertions. ({\bf b}) The $\overline{Z(\beta)}$ path integral with periodic Euclidean time.}
    \end{figure}

The thermal partition function in the Hilbert space defined by the cut $\mathcal{X}$ is
\beq \label{eq:partitionfunction}
Tr_{\mathcal{H}_{\mathcal{X}}}(e^{-\beta H})= \sum_{\gamma} \langle v_\gamma| e^{-\beta H}|v_\gamma\rangle 
= \sum_{\gamma} \zeta[\mathcal{M}_{\gamma}(\beta)] \, .
\eeq
The right side of (\ref{eq:partitionfunction}) is a sum over gravitational path integrals with boundary conditions appropriate to the  survival amplitudes of $\ket{v_\gamma}$ that appear in the middle expression. We call these boundary conditions $\mathcal{M}_{\gamma}(\beta)$. 
By contrast, Gibbons and Hawking \cite{Gibbons:1976ue} proposed, by analogy with quantum field theory, that the  thermal partition function in gravity is computed by the Euclidean path integral with periodic boundary conditions in time
\beq \label{eq:FineTrZ}
\overline{Tr_{\mathcal{H}_{\mathcal{X}}}(e^{-\beta H})}  =\overline{Z(\beta)} \equiv \zeta[\mathcal{M}_{\mathbb{S}_{\beta}\times \mathcal{X}}]
\eeq
where  $\mathbb{S}_{\beta}$ is periodic Euclidean boundary time (Fig.~\ref{fig:BHassump}).
We can also cut the  periodic boundary condition into two ``intervals" $\mathbb{I}_{\beta/2}\times \mathcal{X}$ of length $\beta/2$ to recognize the path integral as the  norm of the state $|\beta \rangle$ defined above: $\overline{Z(\beta)}= \overline{\langle\beta|\beta\rangle}$.  In the Gibbons-Hawking formulation there are no asymptotic matter insertions, unlike in the states $\ket{v_\gamma}$ above which include both gravity and matter.  This leads to a puzzle:
{\it Why is the sum of path integrals in Fig.~\ref{fig:gravpartition} equal to the Gibbons-Hawking path integral in Fig.~\ref{fig:BHassump}?} \footnote{See \cite{Banihashemi:2024weu} for additional discussion on how to put the Gibbons-Hawking path integral on firmer footing. }

\section{The single-sided gravity  partition function}\label{sec:TPF}
We now establish that the Gibbons-Hawking Euclidean gravity path integral with one periodic boundary computes a thermal trace in the Hilbert space.  We will do this in two ways.  First we will show that the result follows on general grounds if the two boundary Hilbert space factorizes into a product of one boundary Hilbert spaces.   Second, we will explicitly perform the thermal trace in a convenient basis and show that it reduces to the Gibbons-Hawking prescription.

\subsection{Hilbert space factorization and the thermal trace}\label{sec:facisGH}
Although the path integral $\overline{Z(\beta)}$ and sum over path integrals $\overline{Tr_{\mathcal{H}_{\mathcal{X}}}(e^{-\beta H})}$ look quite different, we will use
a resolution of the trace and identity to show that they are equal if the two boundary Hilbert space factorizes into a product of one boundary Hilbert spaces, as we have established in \cite{Balasubramanian:2025zey}. Consider the single-boundary gravity Hilbert space $\mathcal{H}_{\mathcal{X}}$ defined by a cut $\mathcal{X}$ with a single connected component, and let $\{\ket{\alpha_i}\}$  denote any (generally non-orthogonal and over-complete) spanning set of states for $\mathcal{H}_{\mathcal{X}}$ made by cutting the path integral. We can resolve the trace in terms of this basis of states by using (\ref{eq:tr}): 
\beq
\overline{Tr_{\mathcal{H}_{\mathcal{X}}}(e^{-\beta H})}=\lim_{n \to -1}\overline{(G^n)_{ij}\langle \alpha_j| e^{-\beta H}|\alpha_i\rangle}=\lim_{n \to -1}\overline{(G^n)_{ij}\langle \alpha_i^*| e^{-\beta H}|\alpha_j^*\rangle},
\eeq
where the last step uses the notation set up in Sec.~\ref{sec:QTFTPI}. Now recall that the path integral for $\overline{Z(\beta)}= \overline{\langle\beta|\beta\rangle}$ can be considered as the overlap of states corresponding to a boundary ``cylinder" of length $\beta /2$. To make contact with this, we split the Hamiltonian evolution by $\beta$ inside the trace into two $\beta /2$ segments and insert the resolution of the identity (\ref{eq:id}) on $\mathcal{H}_{\mathcal{X}}$  in between: 
\beq
\overline{Tr_{\mathcal{H}_{\mathcal{X}}}(e^{-\beta H})}=\overline{Tr_{\mathcal{H}_{\mathcal{X}}}(e^{\frac{-\beta H}{2}}\mathds{1}_{{\mathcal{X}}}e^{\frac{-\beta H}{2}})}=\lim_{n,m \to -1}\overline{(G^n)_{ij}(G^m)_{kl}\langle \alpha_i^*| e^{\frac{-\beta H}{2}}|\alpha_k\rangle \langle \alpha_l|e^{\frac{-\beta H}{2}}|\alpha_j^*\rangle}
\eeq

As explained in Sec.~\ref{sec:QTFTPI}, we can alternatively think of the operator $e^{-\beta H/2}$ as producing the two-boundary state $|\beta\rangle
$, which by construction satisfies
$\langle \alpha_i^*| e^{\frac{-\beta H}{2}}|\alpha_k\rangle=\langle{\beta}\ket{\alpha_i}_L\otimes\ket{\alpha_k}_R$ and similarly $\langle \alpha_l|e^{\frac{-\beta H}{2}}|\alpha_j^*\rangle= \bra{\alpha_j}_L\otimes\bra{\alpha_l}_R|\beta\rangle$. Using these relations we can rewrite the trace in terms of $\ket{\beta}$ and recognize two resolutions of the identity on $\mathcal{H}_{\mathcal{X}}$:
\beq \label{facGHproof}
\overline{Tr_{\mathcal{H}_{\mathcal{X}}}(e^{-\beta H})}= \lim_{n,m \to -1} \overline{(G^n)_{ij}(G^m)_{kl}\langle \beta|\alpha_i\rangle_{L} |\alpha_k\rangle_{R}\langle \alpha_j|_{L}\langle \alpha_l|_{R}|\beta\rangle}= \overline{\langle \beta|\mathds{1}_{{\mathcal{X}_L}} \otimes \mathds{1}_{\mathcal{X}_R}|\beta\rangle}.
\eeq

The state $|\beta\rangle$ is an element of the two-boundary gravity Hilbert space $\mathcal{H}_{\mathcal{X}_L \cup\mathcal{X}_R}$ (see Sec.~\ref{sec:QTFTPI}), while the product $\mathds{1}_{\mathcal{X}_L} \otimes \mathds{1}_{\mathcal{X}_R}$ is the identity on the tensor product of single-boundary theories, $\mathcal{H}_{\mathcal{X}_L}\otimes \mathcal{H}_{\mathcal{X}_R}$.   As discussed in Sec.~\ref{sec:QTFTPI}, \cite{Balasubramanian:2025zey} showed that the two-boundary Hilbert space is actually a product of two single-boundary factors (\ref{eq:facprob}). Hence $\mathds{1}_{\mathcal{X}_L} \otimes \mathds{1}_{\mathcal{X}_R}=\mathds{1}_{\mathcal{X}_L\cup \mathcal{X}_R}$, giving
\beq\label{equality}
\overline{Tr_{\mathcal{H}_{\mathcal{X}}}(e^{-\beta H})}=\overline{\langle \beta|\mathds{1}_{\mathcal{X}_L} \otimes \mathds{1}_{\mathcal{X}_R}|\beta\rangle} \mathbf{=} \overline{\langle \beta| \mathds{1}_{\mathcal{X}_L\cup \mathcal{X}_R}|\beta\rangle} =\overline{\langle \beta|\beta\rangle} \equiv \overline{Z(\beta)} \, .
\eeq
It is easy to repeat  this argument to show \beq \label{eq:squared}
 \overline{\left(Tr_{\mathcal{H}_{\mathcal{X}}}(e^{-\beta H_{\mathcal{X}}})- Z({\beta})\right)^2}{}=0
 \eeq 
 since for any insertion of the trace we can make the above argument.  Together (\ref{facGHproof}) and (\ref{eq:squared}) show that
 \beq \label{finegraintrZ}
 Tr_{\mathcal{H}_{\mathcal{X}}}(e^{-\beta H})= Z(\beta)\, ,
 \eeq
 as a fine-grained equality.

Remarkably, the  argument we have given does not require an explicit construction of the basis of states, or any particular form of the gravitational action. The argument instead relies on two requirements: ({\bf 1}) The gravity Hamiltonian must generate asymptotic boundary time evolution; ({\bf 2}) The Hilbert space of  gravity  with two asymptotic Lorentzian boundaries must factorize into copies of the single-boundary Hilbert space. We emphasise that, once these two criteria are granted, the chain of equalities~(\ref{equality}) is purely linear-algebraic: no gravitational input beyond the factorisation result enters. The role of gravity is confined to establishing $\mathds{1}_{\mathcal{X}_L}\otimes\mathds{1}_{\mathcal{X}_R}=\mathds{1}_{\mathcal{X}_L\cup\mathcal{X}_R}$, which is the content of \cite{Balasubramanian:2025zey}.

Hence in any theory of quantum gravity with an (effective) path integral formulation satisfying these conditions, the Gibbons-Hawking prescription for the thermal partition function \textit{is} in-fact tracing over the states in the theory. The first requirement seems universally true in gravity theories with a time-like asymptotic boundary as it is a direct consequence of invariance under local bulk diffeomorphisms, which any gravitating system must satisfy by definition. Factorization of the Hilbert space is however non-trivial. To derive it, \cite{Balasubramanian:2025zey} had to include the effects of  nonperturbative wormholes  connecting the asymptotic boundaries used to prepare two-boundary and single-boundary gravity states. By contrast, the perturbative Hilbert space  does not factorize on its own, suggesting that  the Euclidean gravity path integral with periodic time does not equal the thermal trace in perturbative gravity.  The sum over topologies in non-perturbative gravity is essential.

We stress at the outset that the factorisation argument of
Sec.~\ref{sec:facisGH} already establishes $Tr_{\mathcal{H}_{\mathcal{X}}}
(e^{-\beta H})=Z(\beta)$ as a statement about the "full" path integral:\footnote{We use ``full path integral'' with a caveat. In $d>2$ the Euclidean
gravitational path integral is not known to be well-defined nonperturbatively
--- the conformal mode renders the action unbounded below, and the sum over
topologies is at best an asymptotic series --- so we mean by it the formal
object defined by the bulk effective theory, organised as a sum over geometries
and topologies, without presuming that this object converges or admits a unique
completion. The factorisation argument does not require that it does: it reduces
the equality to the operator identity $\mathds{1}_{\mathcal{X}_L}\otimes
\mathds{1}_{\mathcal{X}_R}=\mathds{1}_{\mathcal{X}_L\cup\mathcal{X}_R}$
established in \cite{Balasubramanian:2025zey}, so the equality holds at whatever
level that identity does --- equivalently, in any nonperturbative completion in
which the two-boundary Hilbert space factorises --- rather than relying on
convergence of the sum. This is the sense intended by ``full path integral
equality'': it survives the inclusion of arbitrary, possibly off-shell,
geometries compatible with the boundary conditions (cf.\ Sec.~5 of
\cite{Balasubramanian:2025jeu}).} once the two-boundary Hilbert space factorises, the remaining manipulations are
purely algebraic, so the equality holds in exactly the same sense, and to
exactly the same extent, as the factorisation identity itself, with no reference
to a saddle expansion. The explicit evaluation in
this section is therefore best read as an independent and more concrete check
of that conclusion, one that operates mostly at the level of the sum over
saddles.

\subsection{Wormholes, a basis, and the trace} \label{eq:directcal}
\begin{figure}[h]
    \centering
    \begin{subfigure}{0.45\linewidth}
    \centering
    \includegraphics[width=0.3\linewidth]{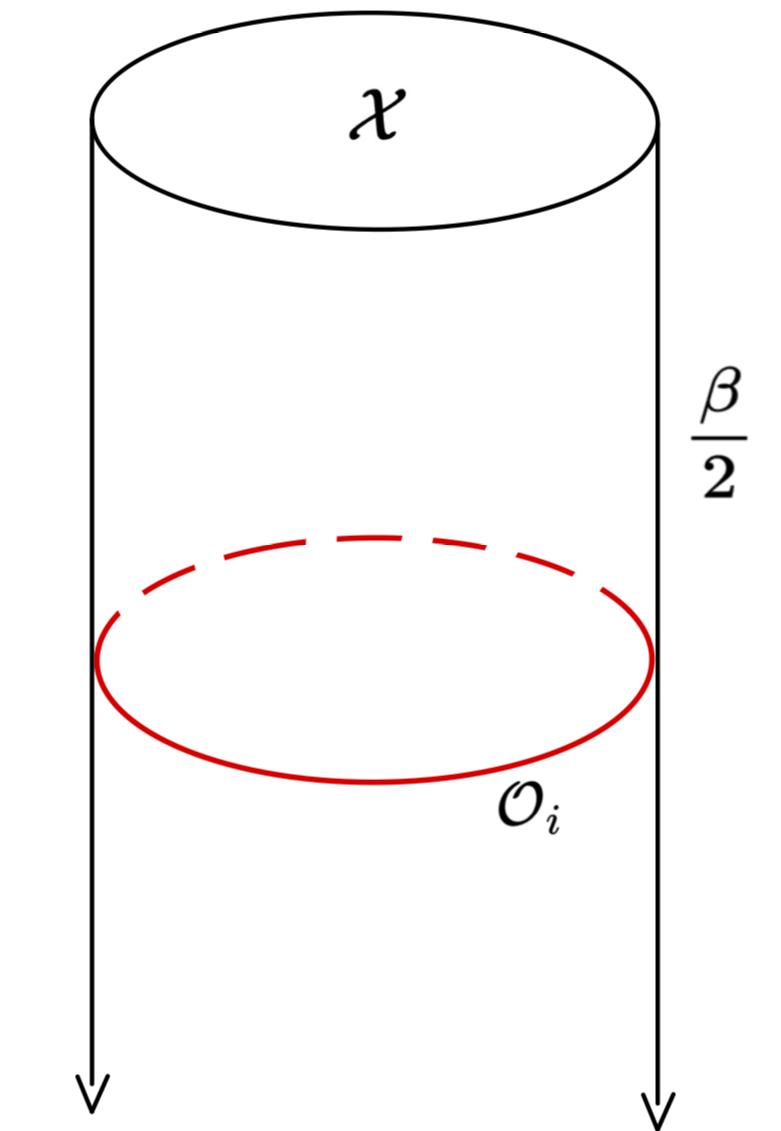}
    \caption{}
    \end{subfigure}
     \centering
    \begin{subfigure}{0.45\linewidth}
    \centering
    \includegraphics[width=0.3\linewidth]{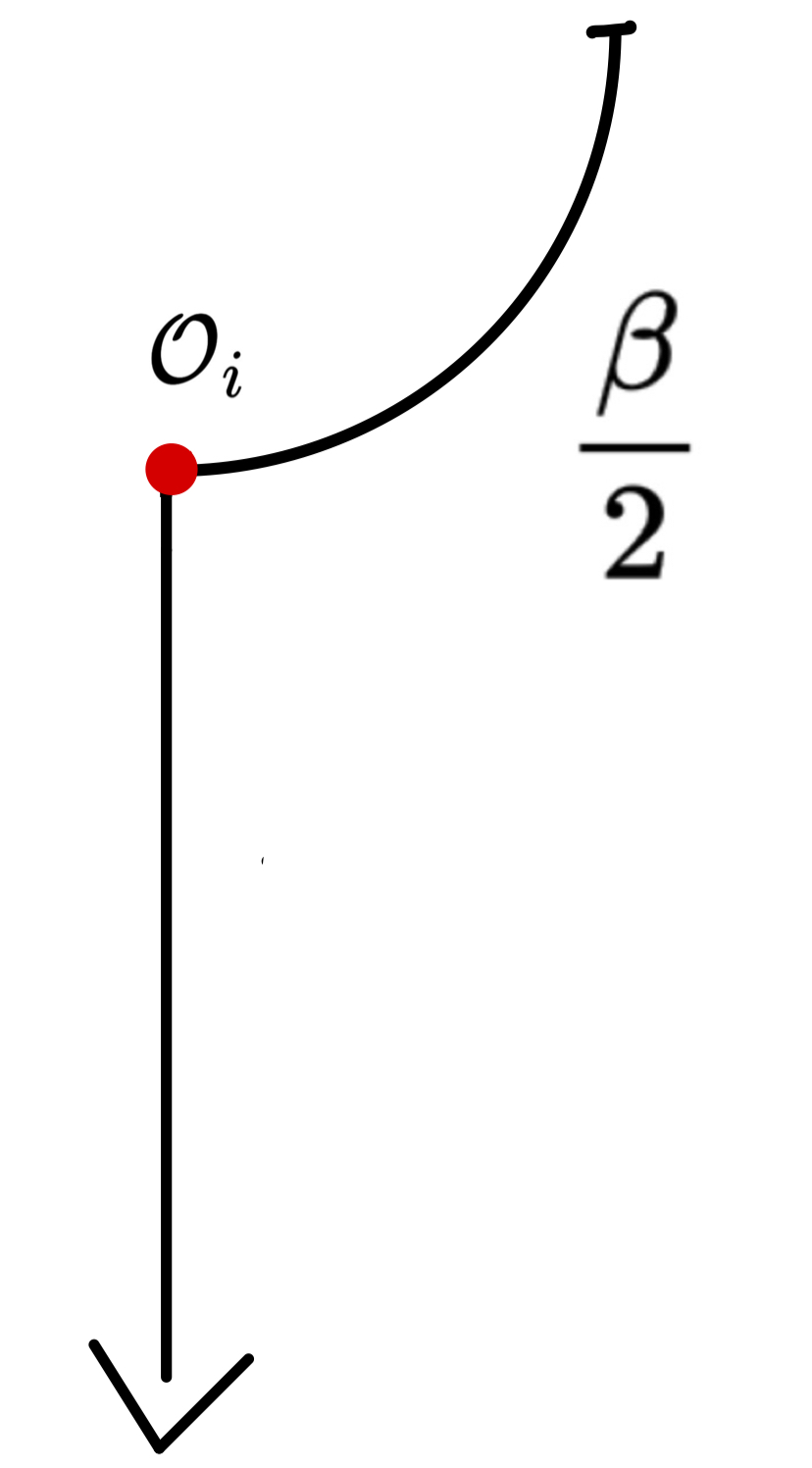}
    \caption{}
    \end{subfigure}

    \caption{Path integral boundary condition defining the single-sided shell states.  ({\bf a}) Euclidean boundary with topology $\mathbb{R}^{<0}\times\mathbb{S}^{d-1}$ for preparation of the shell states.  The arrows indicate a half-infinite line. 
    The shell operator $\mathcal{O}_{i}$ is pictured in red and $\beta_{}/2$ is the Euclidean ``preparation time''. ({\bf b}) Euclidean boundary with the $\mathbb{S}^{d-1}$ suppressed. We adopt this convention, and sometimes depict this boundary with a curve or a kink to clarify diagrams (images adapted from \cite{Balasubramanian:2025zey}).}

     \label{fig:sinlge_def}
\end{figure}

As an alternative approach, we will explicitly perform the thermal trace in  the ``shell state" basis for the single boundary Hilbert space \cite{Chandra:2022fwi,Balasubramanian:2025zey}.   To define these states we fix a boundary with topology $\mathbb{R}^{<0}\times \mathbb{S}^{d-1}$ (where $\mathbb{R}^{<0}$ is the half line) and insert an $\mathbb{S}^{d-1}$ symmetric heavy ($m \sim \mathit{O}(1/G_{N})$) dust shell operator  $\mathcal{O}_{S}$ separated by a boundary length $\frac{\beta}{2}$ from the cut $\mathcal{X}$.  Thus, the boundary time extends from $-\infty$ to  $\tau = 0$,  where $\mathcal{O}_S$ is inserted, followed by a  time section up to $ \tau =\beta/2$ (Fig.~\ref{fig:sinlge_def}).  We can obtain an infinite family of shell states in this way by varying the mass $m_i$ of the  operators $\mathcal{O}_{i}$ to produce states  $|i\rangle$.

\begin{figure}
    \centering
    \includegraphics[width=0.5\linewidth]{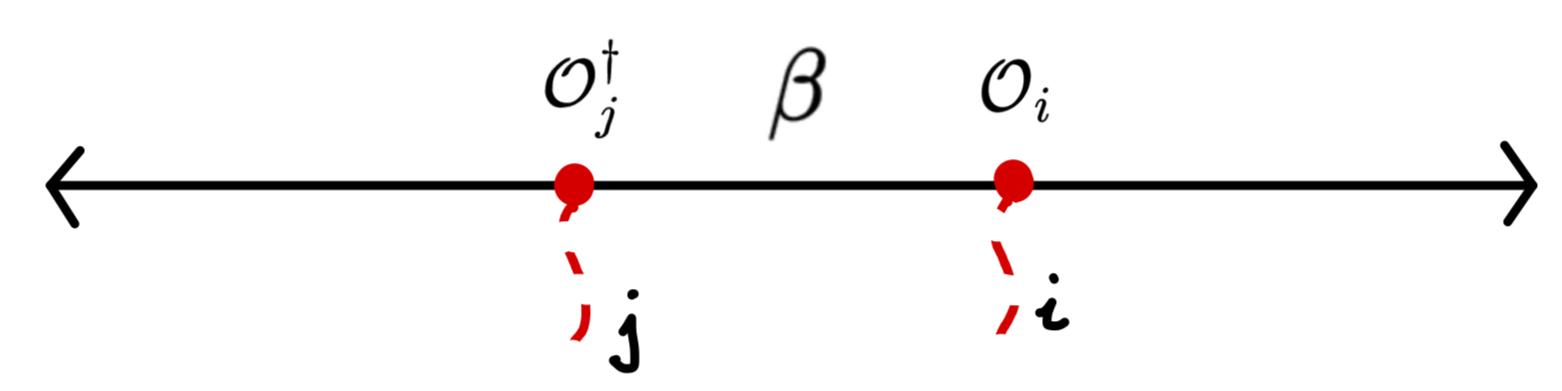}
    \caption{Shell-strip asymptotic boundary condition for the overlap $\braket{j|i}$ consisting of the line $\lim_{\alpha \to \infty} [-\alpha, \beta + \alpha]$ on which  $\mathcal{O}_{i}$ and  $\mathcal{O}^{\dagger}_{j}$ are inserted at $\tau =0 $ and  $\tau=\beta$ respectively (images adapted from \cite{Balasubramanian:2025zey}).}
    \label{fig:1s_shellbdry}
\end{figure}
\begin{figure}
    \centering
    \includegraphics[width=0.8\linewidth]{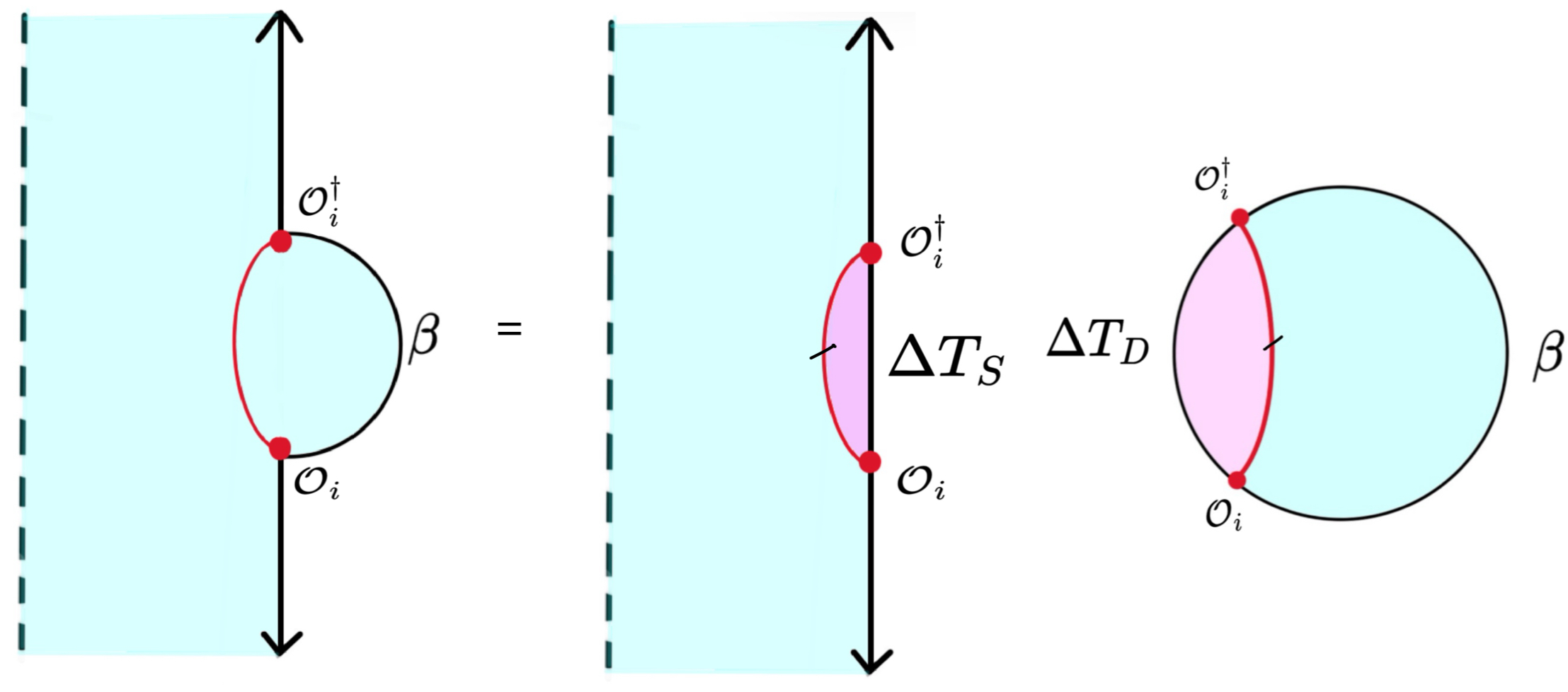}
    \caption{The saddle for the norm $\overline{\braket{i|i}}$ is constructed by considering the shell propagating on a disk and strip separately for some propagation times $\Delta T_{S,D}$ and then gluing them together along the $i$-shell worldvolume by discarding the shell homology region (purple). The junction conditions dynamically determine $\Delta T_{S,D}$ to yield an on shell glued geometry. We have suppressed the angular directions in these diagrams, and represent the radius-time plane with the dashed line representing the origin (images adapted from \cite{Balasubramanian:2025zey}).}
    \label{fig:singe_norm}
\end{figure}

We will need to consider path integrals with a boundary   consisting of a Euclidean half line $[-\infty,0]$ followed  operator insertions and Euclidean time evolution up to  time $T_{E}$, and then another  half line $[T_{E},\infty]$.\footnote{See \cite{Balasubramanian:2025zey} and Fig.~\ref{fig:1s_shellbdry} for a specification of the limit procedure in which these strip boundary conditions are defined. We do not specify the details here as we will not need them.} We will call $T_{E}$  the ``length of the strip'' and denote the path integral with this  boundary condition as $S(T_{E})$.  For example, to compute the  Gram matrix element $G_{ij}\equiv \langle i|j\rangle$ we construct a ket  by the path integral  in Fig.~\ref{fig:sinlge_def}, and a bra  defined by a similar path integral on  $\frac{\beta}{2}<\tau<\infty$.  The boundary condition defining $G_{ij}$ is thus an interval $[-\infty,0]$ followed by $\mathcal{O}_j$ insertion, boundary time evolution by an amount  $\beta$, $\mathcal{O}^{\dagger}_i$ insertion, and another  half line $[\beta,\infty]$ (Fig.~\ref{fig:1s_shellbdry}).   In the above terminology this is a strip of length $\beta$ containing  $\mathcal{O}_{j}$ and  $\mathcal{O}^{\dagger}_{i}$ operator insertions.  The coarse-grained overlap $\overline{\langle i|j \rangle}$ is   evaluated by the path integral with such a {\it  shell strip boundary condition }(Fig.~\ref{fig:singe_norm}).  Following~\cite{Balasubramanian:2022gmo},  we  make different shell states orthogonal, $\overline{\langle i|j \rangle} =\delta_{ij} Z_{1} $, by taking their inertial mass differences to be arbitrarily large, since  it will take  $|m_{i}-m_{j}|$ bulk interactions in Planck units for such shells to annihilate.

\paragraph{Norm.}\label{sec:appStripNorm}  As shown in \cite{Balasubramanian:2025zey}, we can construct saddlepoints for path integrals with the  shell strip boundary condition by filling in one side of the shell worldvolume with any saddle geometry of $\overline{Z(\beta)} = \overline{\braket{\beta|\beta}}$  (the ``disk'') and the other with a Euclidean vacuum geometry with non-compact time (the ``strip"). We use the Israel junction conditions \cite{Israel:1966rt} to glue the geometries at the shells  (Fig.~\ref{fig:singe_norm}). For example, suppose we take a Euclidean AdS strip with a shell operator  $\mathcal{O}_{i}$ on the boundary at $\tau=0$ and the conjugate operator $\mathcal{O}^{\dagger}_{i}$  at time $\tau=\Delta T_{S}$ and a Euclidean AdS disk with boundary insertions $\mathcal{O}_{i}$ and $\mathcal{O}^{\dagger}_{i}$ separated by  $\Delta T_{D}$ on a disk of boundary circumference $\beta +\Delta T_{D}$.  To construct the saddlepoint geometry we  discard the shell homology regions (purple  in Fig.~\ref{fig:singe_norm}) from the strip and disk and use the junction conditions to glue  the shell world-volumes. The junction conditions dynamically determine $\Delta T_{S}$ and $\Delta T_{D}$ such that the resulting geometry satisfies the equations of motion.

\paragraph{Large shell mass limit.} 
As discussed in \cite{Balasubramanian:2025zey}, as $m_i \to \infty$ limit the  turning point of the shell approaches the asymptotic boundary and $\Delta T_{S},\Delta T_{D} \to 0$. So the shell homology regions pinch off and the remainder grows to cover the entire strip and disk.  In this limit, the shell contribution to the action is a universal factor $Z_{m_i}$ and the overlap is:
\beq \label{eq:overlap}
\overline{\langle i|j \rangle} = \delta_{ij} \, Z_{m_i} \times 
\overline{Z}(\beta)\times \overline{S(}0)\, .
\eeq
 As the above construction could have been done using any of the saddles for $\overline{Z(\beta)}$, (\ref{eq:overlap}) equates the sum over saddlepoints of the LHS to that of the RHS.

\begin{figure}[h]
\begin{subfigure}[c]{0.48\linewidth}
        \centering
        \includegraphics[width=\linewidth]{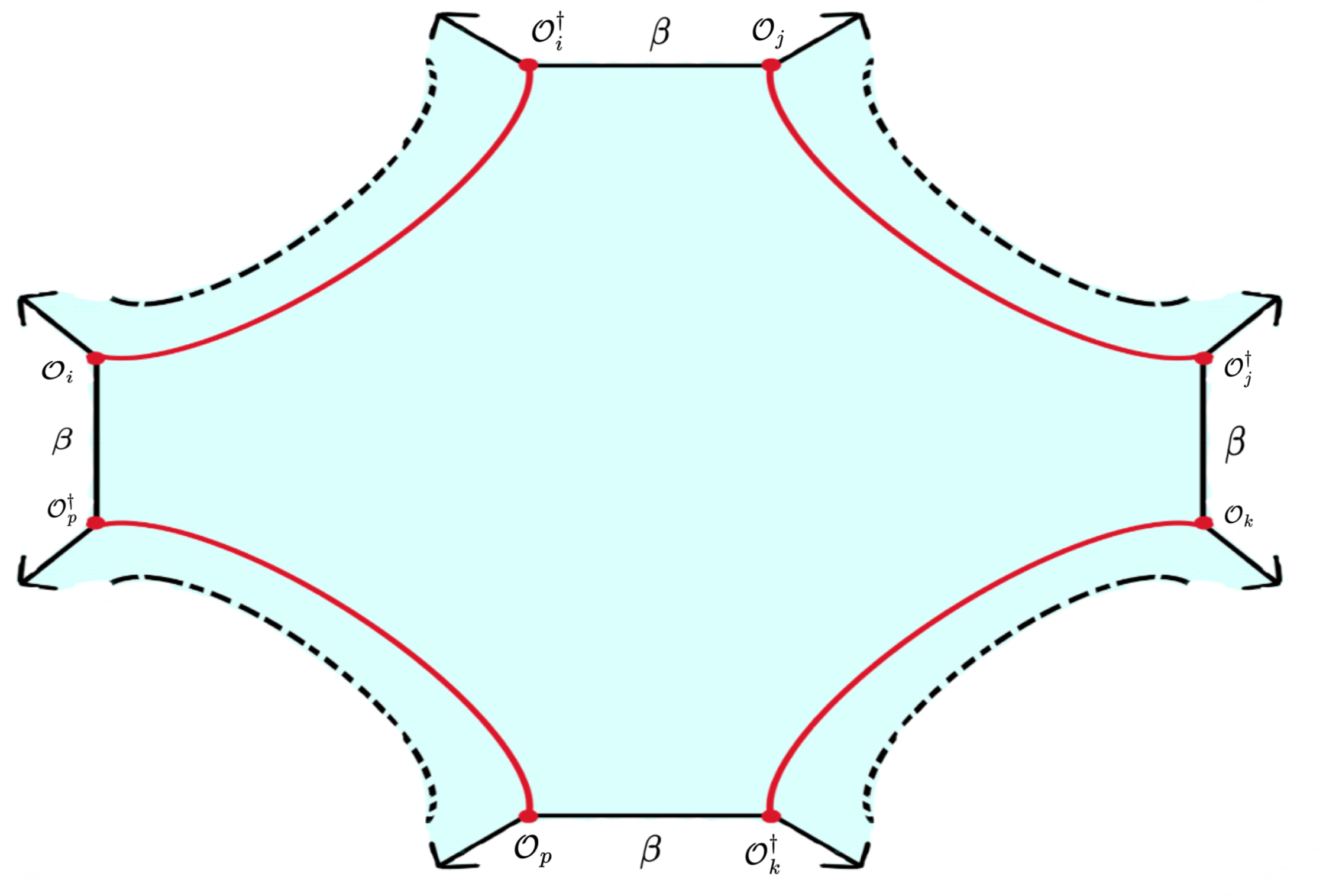}
        \caption{}
        \label{fig:Single_wormhole}
\end{subfigure}
\hfill
\begin{subfigure}[c]{0.48\linewidth}
    \centering
    \includegraphics[width=\linewidth]{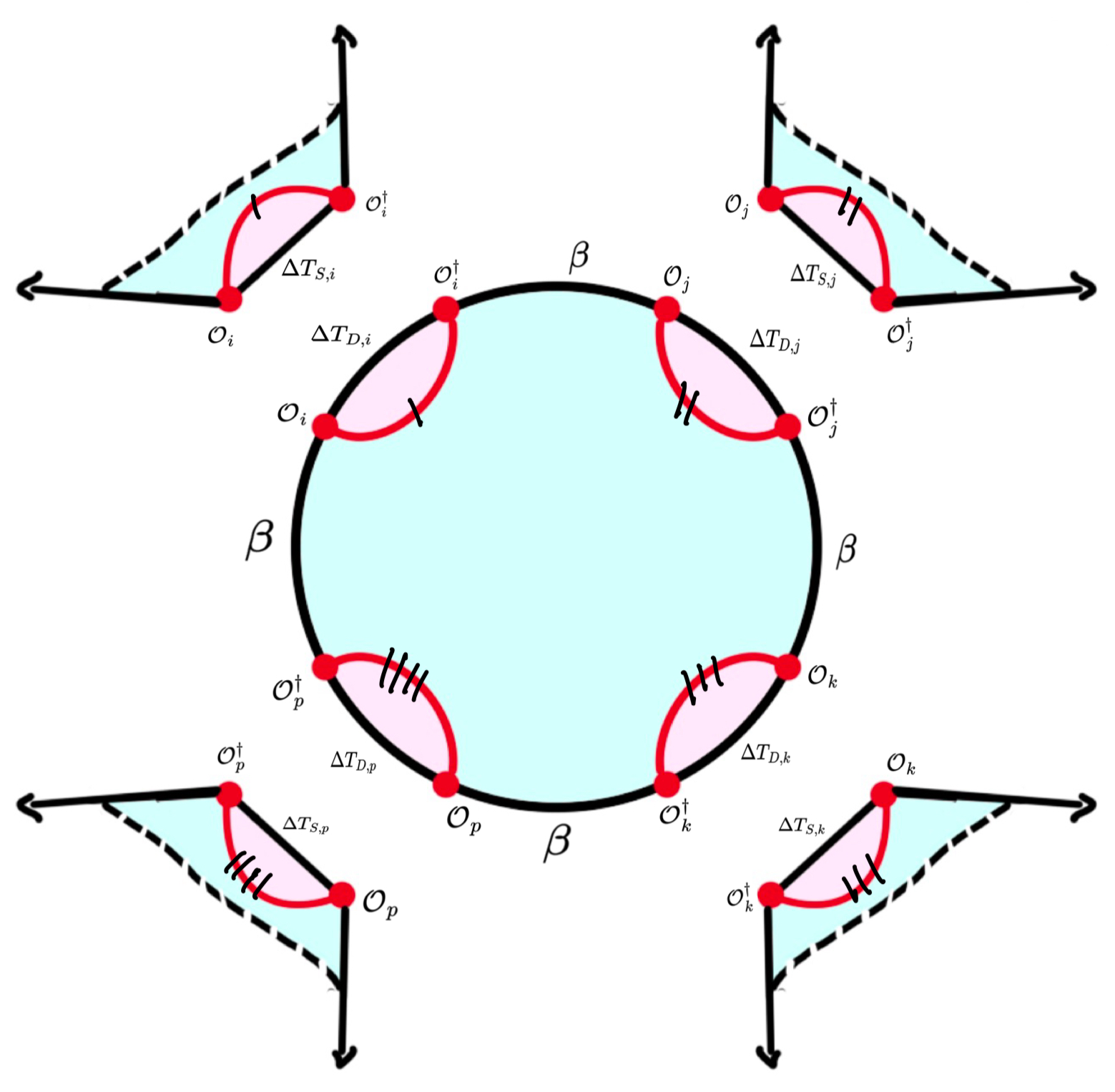}
    \caption{}
    \label{fig:planarWHconstruct}
\end{subfigure}
\caption{The fully connected single-sided shell wormhole saddles for $\overline{(G^n)_{ii}}$ are constructed by gluing $n$ strips into the disk along the shell worldvolumes, depicted here for $n=4$. Images adapted from \cite{Balasubramanian:2025zey}.}
\end{figure}

\paragraph{Wormholes.} \label{sec:1sWH} 
To construct fully connected wormhole saddles contributing to \beq \label{eq:fullconn}
\overline{(G^n)_{ii}}=\overline{\braket{i|j_1}\braket{j_1|j_2} \cdots \braket{j_n|i}} \eeq
(no sum on $i,j_k$) we connect  $n$  boundaries (from each factor in $(G^n)_{ii}$) through a single  wormhole. The $k$-th shell must propagate from $\mathcal{O}_{k}$ to $\mathcal{O}^{\dagger}_{k}$ and so the wormhole connectivity reflects the index structure of $\overline{(G^n)_{ii}}$ (Fig.~\ref{fig:Single_wormhole}).  Similarly to the norm, we construct  the  saddle by gluing  $n$ copies of the Euclidean strip into a disk along the shell word-volumes  (Fig.~\ref{fig:planarWHconstruct}).  To construct saddle geometries satisfying this boundary condition consider a disk with $n$ shells inserted on the asymptotic boundary, $\mathcal{O}_{S_i}$ separated from $\mathcal{O}^{\dagger}_{S_i}$ by boundary time $\Delta T_{D,i}$, and $\mathcal{O}^{\dagger}_{S_i}$ from $\mathcal{O}_{S_{i+1}}$ by boundary time  $\beta$ in a circular pattern. Separate shell-strips are constructed by placing a shell $\mathcal{O}_{S_i}$ on the strip boundary separated from $\mathcal{O}^{\dagger}_{S_i}$ by $\Delta T_{S,i}$ boundary time. We then remove the shell homology regions on the disk and strips and use the junction conditions to identify corresponding shell worldvolumes, and to dynamically determine $\Delta T_{D,i}$ and $\Delta T_{S,i}$.  Details are  in \cite{Balasubramanian:2025zey}. As described above, the shell propagation times go to  zero in the large shell-mass limit, so the saddle  becomes  a product of  $n$ strips, a  disk of length $n\beta$, and   universal  shell contributions: $\overline{(G^{n})_{ii}}= \overline{Z}(n\beta) \times \overline{S}(0)^n \times \prod_{i=1}^n Z_{m_i}$. Normalizing  shell states to have unit norm
 we get 

\beq \label{eq:1sshellwh}
\ket{k}\rightarrow \frac{\ket{k}}{\sqrt{Z_{m_k} \times 
\overline{Z(}\beta) \times \overline{S}(0)}} ~~~~~;~~~~~
\overline{(G^{n})_{ii}}= \frac{\overline{Z}(n\beta)}{\overline{Z}(\beta)^n} \, .
\eeq

\paragraph{Saddle geometries.} 
Following \cite{Balasubramanian:2025zey}, to associate geometry to a shell state we consider the leading gravitational saddlepoint for 
the norm $\overline{\langle i|i\rangle}$. In the large shell mass limit resulting in (\ref{eq:overlap}), this is the determined by the leading saddle for $\overline{Z(\beta)}$. In the asymptotically AdS case the leading saddle is either thermal Euclidean AdS if $\beta > \beta _{HP}$  or the large Euclidean  black hole if $\beta < \beta _{HP}$, where $\beta_{HP}$ marks the Hawking-Page transition \cite{Hawking:1982dh}. With asymptotically flat boundary conditions,  thermal flat space is always leading, although after a micro-canonical projection the black hole saddle can be leading. The time reflection symmetric  slice of these saddles is  continued to Lorentzian signature. There are thus two classes of single sided shell states:
\begin{itemize}
\item {\bf Type A}: The leading saddle is the Euclidean  black hole, so the Lorentzian geometry is a black hole with a shell  in the interior, capped off by the vacuum behind the shell. We will refer to these  as \textit{type A} states. In AdS space we find these saddles when  $\beta < \beta _{HP}$ and in asymptotically flat space in the finite energy micro-canonical ensemble. 
\item  {\bf Type B}: The leading saddle has a non-contractible thermal circle  and the Lorentzian geometry is the thermal space with a disconnected  closed ``Big Crunch" universe in which the shell propagates. We will refer to these states as \textit{type B} shell states. In AdS space we find these saddles for  $\beta >\beta _{HP}$ and in asymptotically flat space for any $\beta $ in the canonical ensemble. 
\end{itemize}

\begin{figure}[h]
  \begin{subfigure}[c]{0.42\linewidth}
    \centering
    \includegraphics[width=0.7\linewidth]{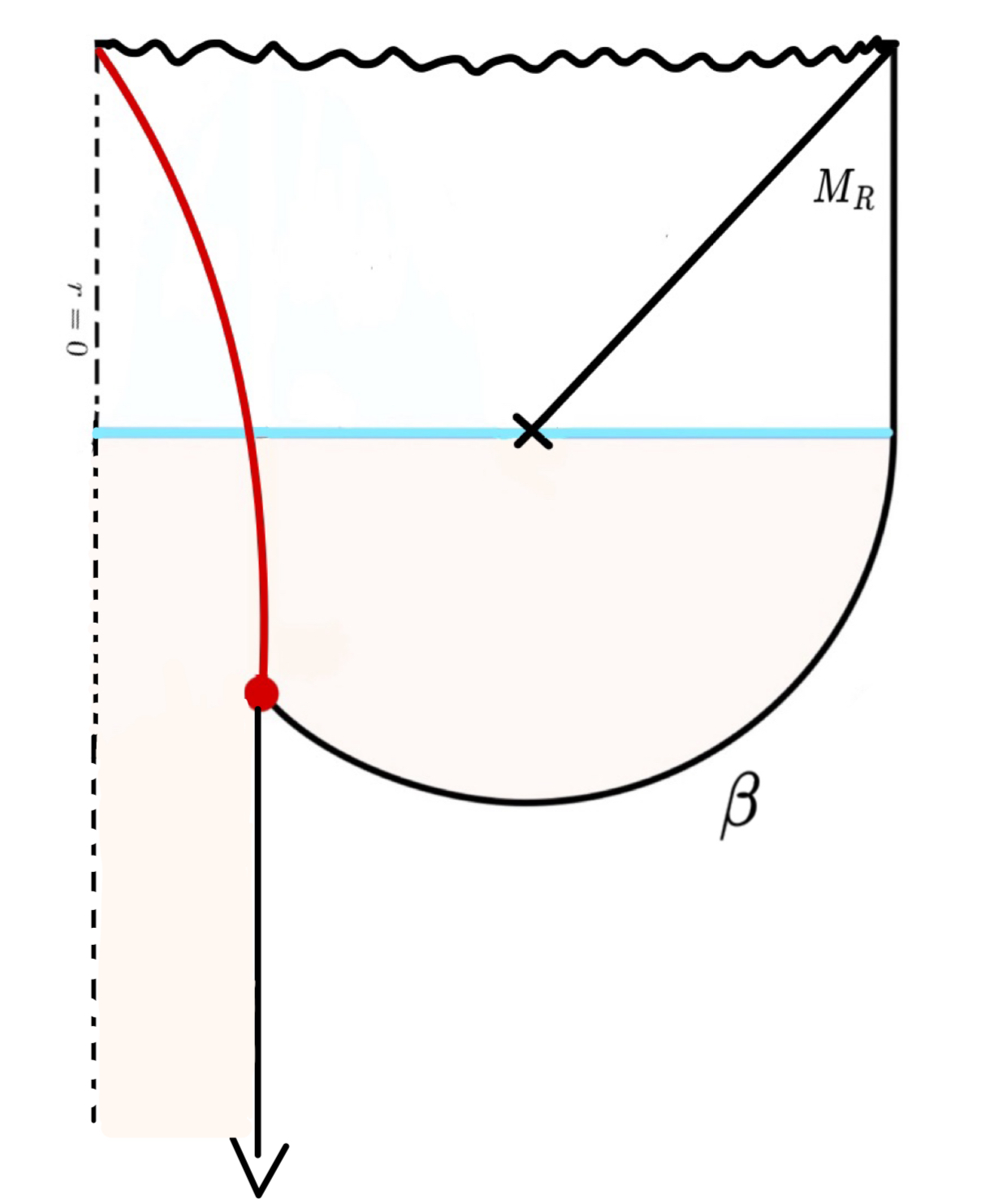}
    \caption{}
    \label{fig:type1}
    \end{subfigure}
    \hfill
 \begin{subfigure}[c]{0.42\linewidth}
    \centering
    \includegraphics[width=0.7\linewidth]{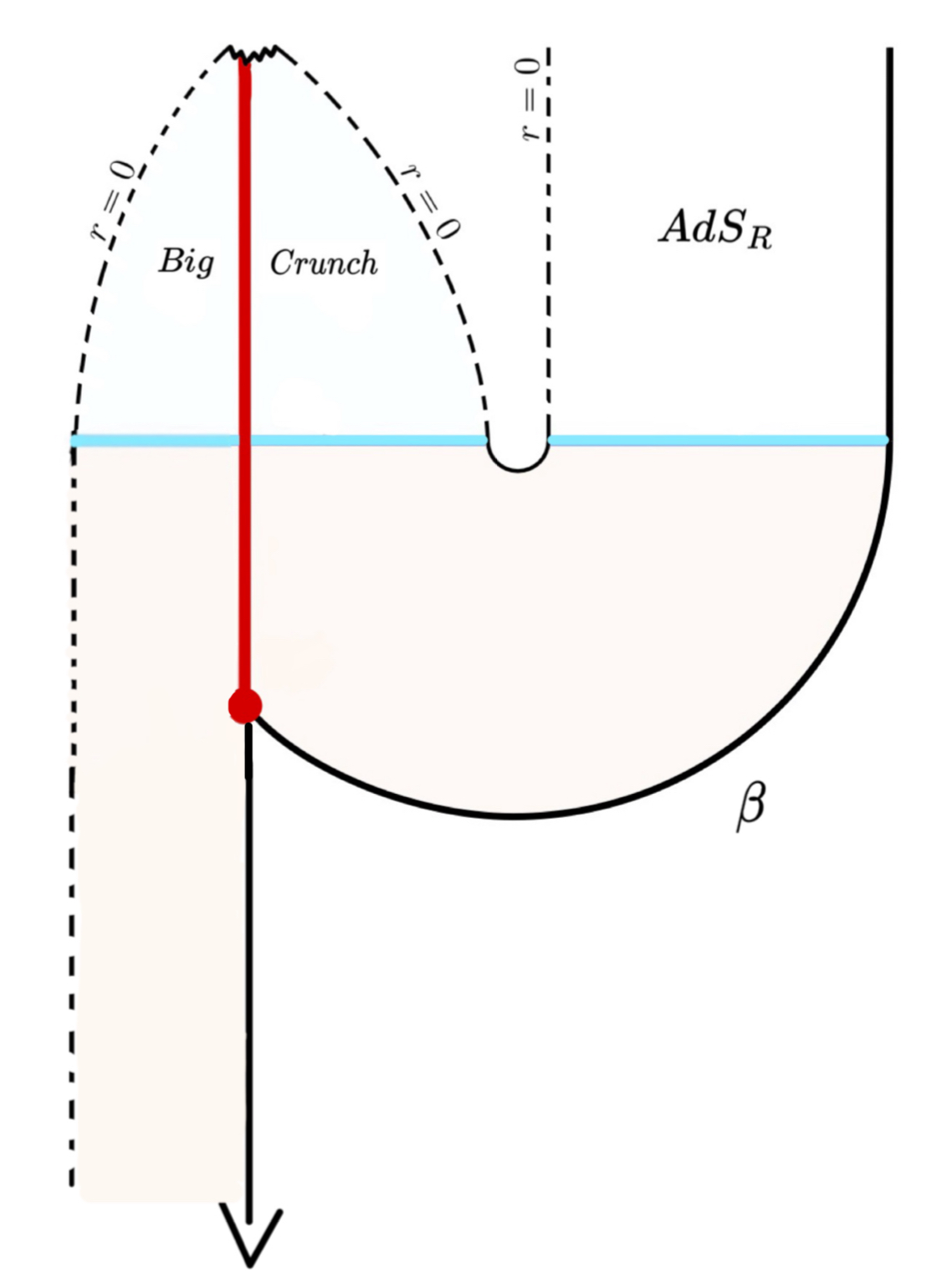}
    \caption{}
    \label{fig:type2}
    \end{subfigure}
\caption{Analytic continuation of the single-sided shell state saddlepoints to  Lorentzian signature, shown here with asymptotically AdS boundary conditions. For asymptotically flat boundary conditions the vertical lines are replaced by diamonds.  ({\bf a}) Type A shell state corresponding to a single-sided black hole. ({\bf b}) Type B shell state consisting of thermal AdS (or thermal flat space)  with and added disconnected compact Big-Crunch AdS cosmology. Images adapted from \cite{Balasubramanian:2025zey}. }
\end{figure}

\subsubsection{Explicit trace}
Let $\mathcal{H}_{\mathcal{X}}$ be the single boundary Hilbert space. To establish the fine-grained 
equality $\mathcal{A}=Tr_{\mathcal{H}_{\mathcal{X}}}(e^{-\beta H_{\mathcal{X}}}) = Z(\beta) = 
\mathcal{B}$, we will use the toolkit from \cite{Balasubramanian:2025jeu} (see Sec.~\ref{sec:QTFTPI}) to 
show that  $\overline{\mathcal{A}-\mathcal{B}}=0$ and $\overline{(\mathcal{A}-\mathcal{B})^2}=0$ 
directly in the single boundary theory.  Denote a single-sided shell state of mass $m_i$ and preparation 
temperature $\beta_R$ as $|i\rangle_{R}$, and the Hilbert space spanned by $\kappa_R$ such states of 
varying mass but fixed preparation temperature as $\mathcal{H}_R$.  The single-sided shell states span 
$\mathcal{H}_{\mathcal{X}}$ as $\kappa_R \to \infty$ \cite{Balasubramanian:2025zey}  (we return below to
the weaker condition $\kappa_R\gtrsim e^{S}$ that suffices in the
micro-canonical ensemble). Hence the trace over $\mathcal{H}_{\mathcal{X}}$ can be resolved as: 
\beq \label{eq:1sTrisZ}
\overline{Tr_{\mathcal{H}_{\mathcal{X}}}(e^{-\beta H_{\mathcal{X}}})} = \overline{Tr_{\mathcal{H}_{R}}(e^{-\beta H_{\mathcal{X}}})}= \lim_{n \to -1} \overline{(G^{n})_{ij} \langle j|e^{-\beta H_{\mathcal{X}}} |i \rangle} \, .
\eeq
where we are summing over $i,j$.

\begin{figure}
\centering
\begin{subfigure}[c]{0.45\linewidth}
    \centering
    \includegraphics[width=\linewidth]{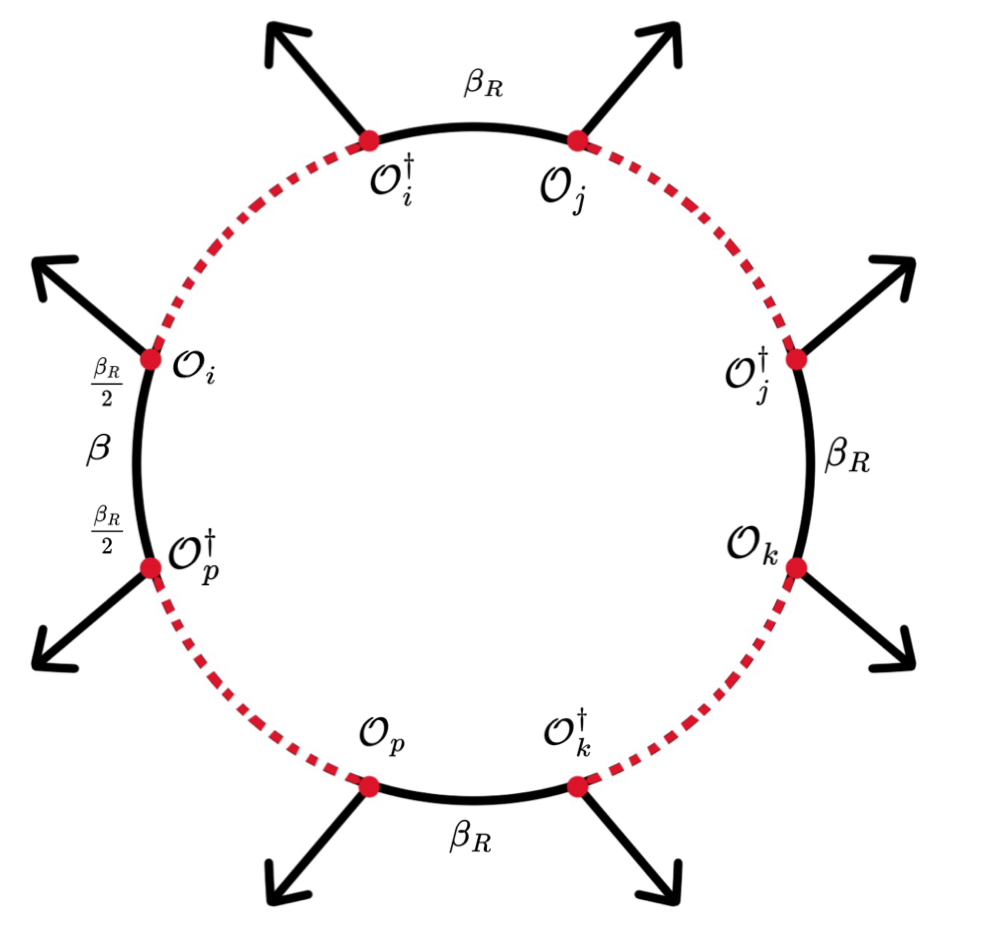}
    \caption{}
    \label{fig:1sTRBCa}
\end{subfigure}
    \begin{subfigure}[c]{0.45\linewidth}
    \centering
    \includegraphics[width=\linewidth]{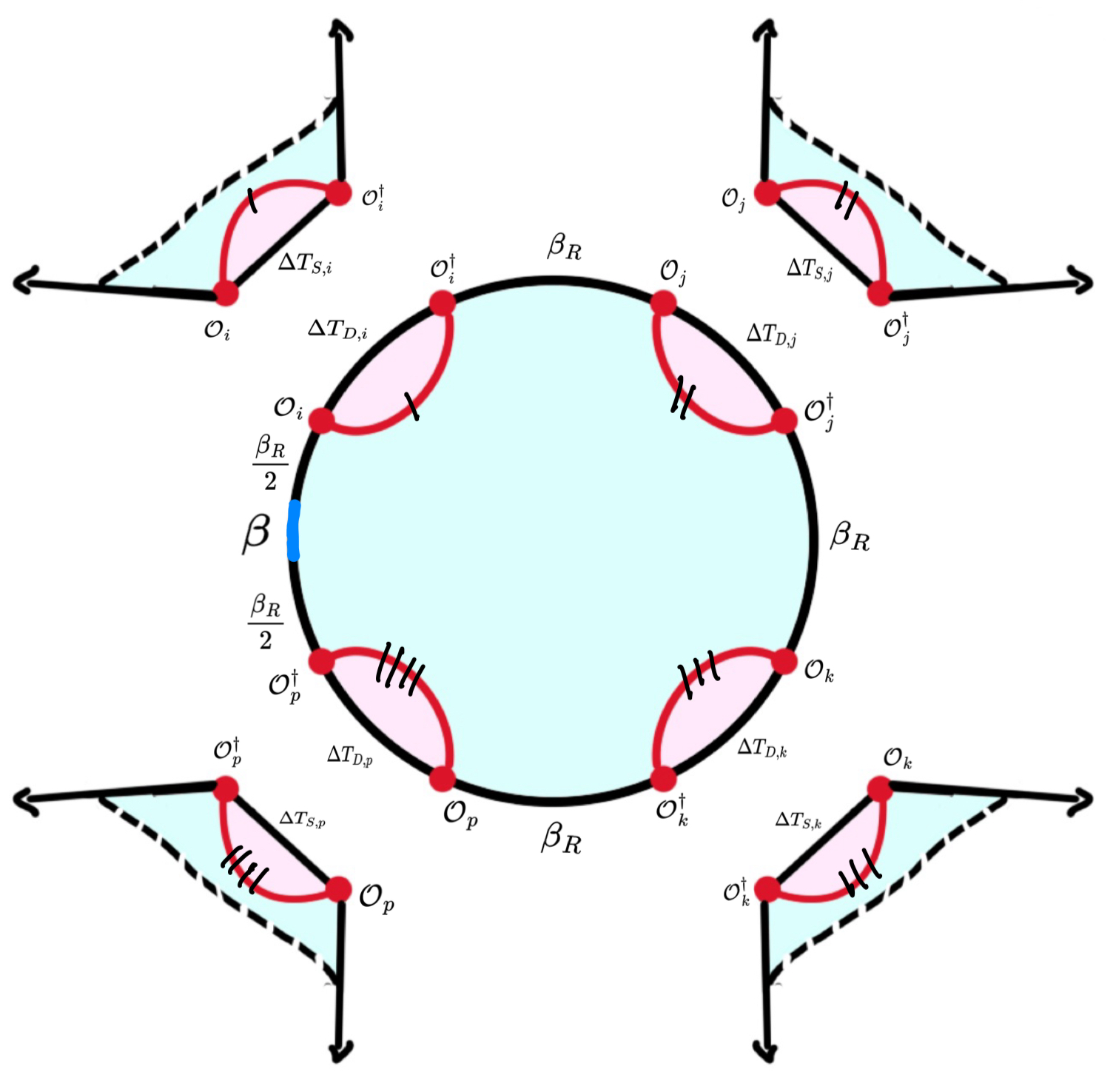}
    \caption{}
    \label{fig:TrSCWHb}
    \end{subfigure}
    \caption{({\bf a}) Boundary condition for the path integal computing (\ref{eq:1sTrisZ}) with $n=3$. ({\bf b}) Construction of the semiclassical saddle for (\ref{eq:1sTrisZ}) by gluing shell-strips into a disk.}
\end{figure}

Consider the saddlepoint approximation of $\overline{(G^{n})_{ij} \langle j|e^{-\beta H_{\mathcal{X}}} |i \rangle}$.    As discussed above, powers of the single-sided Gram matrix contribute shell-strip insertions of length $\beta_R$ (Fig.~\ref{fig:1s_shellbdry}), while  $\langle j|e^{-\beta H_{\mathcal{X}}} |i \rangle$ contributes  to a lengthened shell strip boundary of length $\beta + \beta_R$.  To evaluate the gravitational path integral we should sum over  topologies. However,  in the $\kappa_R \to \infty$ limit in which the shell states span  $\mathcal{H}_{\mathcal{X}}$, only geometries that do not break any shell index loops contribute \cite{Balasubramanian:2025jeu}.  For (\ref{eq:1sTrisZ}) this means that contributing  geometries  fill in the boundary index loop (Fig.~\ref{fig:1sTRBCa}) with a single fully-connected bulk. 
 This  loop structure of the trace will lead to the  circular boundary condition defining $Z(\beta)$.

These fully-connected saddle geometries are the same as those constructed for (\ref{eq:fullconn}) above, except that  the  $\mathcal{O}^{\dagger}_{S_{n+1}}$ and $\mathcal{O}_{S_1}$ shell insertions on the boundary are separated by $\beta + \beta_R$ because of the additional $e^{-\beta H_{\mathcal{X}}}$ time evolution. So, by the same logic, in the large shell mass limit the trace equals $\overline{Z}(\beta + (n+1) \beta_R) \times \overline{S}(0)^{n+1}\times \prod_{i=1}^{n+1}Z_{m_i}$, and after normalizing the shell states as in (\ref{eq:1sshellwh}) yields
\beq \label{sctrace}
\overline{(G^{n})_{ij}\langle 
j|e^{-H_R\beta}|i\rangle} = \kappa_R^{n+1} \frac{\overline{Z}(\beta + (n+1) \beta_R) }{Z(\beta_R)^{n+1}} \, .
\eeq
Hence in the $n \to -1$ limit
\beq \label{Tr=Z}
\overline{Tr_{\mathcal{H}_{\mathcal{X}}}(e^{-\beta H_{\mathcal{X}}})}= \overline{Z(\beta)}\, . 
\eeq
This equality includes all the saddlepoints contributing to the left and right sides.

As explained in \cite{Balasubramanian:2025jeu}, in the $\kappa_R \to \infty$ limit we can readily repeat this argument to also show $\overline{\left(Tr_{\mathcal{H}_{\mathcal{X}}}(e^{-\beta H_{\mathcal{X}}})- Z({\beta})\right)^2}{}=0$ establishing the fine-grained equality $Tr_{\mathcal{H}_{\mathcal{X}}}(e^{-\beta H})= Z(\beta)$ in the saddlepoint approximation. Once again, the relation (\ref{Tr=Z}) is only true in the non-perturbative theory, and would fail in a perturbative description that did not include wormhole contributions to the path integral.

\paragraph{Matching the full path integral.}
The equality~(\ref{Tr=Z}) extends from the leading saddles to the full
path integral. The cleanest statement of this is the factorisation argument of
Sec.~\ref{sec:facisGH}: there, $Tr_{\mathcal{H}_{\mathcal{X}}}(e^{-\beta H})=
\langle\beta|\mathds{1}_{\mathcal{X}_L}\otimes\mathds{1}_{\mathcal{X}_R}|\beta
\rangle=\langle\beta|\beta\rangle=Z(\beta)$ followed purely algebraically once
the two-boundary Hilbert space factorises, and is therefore valid for the full
path integral whenever the latter is suitably defined, independently of any
saddle expansion. The explicit shell-basis computation of this section provides
an independent check of the same statement, largely at the level of the sum over
saddles. It can be extended beyond the leading saddle by a heuristic argument,
which we sketch but do not aim to make precise, as follows.
First, the saddle geometries of the two sides are in one-to-one correspondence.
The shell strip admits a single saddle, while the central disk into which the
strips are glued admits one saddle for each saddle of the Gibbons-Hawking path
integral $\overline{Z(\beta)}$ --- these disk geometries are precisely the
$\overline{Z(\beta)}$ saddles being glued in. Hence every saddle contributing to
the explicit trace corresponds to exactly one saddle of $\overline{Z(\beta)}$,
and vice versa.
Second, the perturbative corrections around each pair of corresponding saddles
match. In the large shell-mass limit each strip decouples into an independent
region carrying its own perturbative expansion (one may, for instance, dress
each strip with its own loop corrections), and the heavy shells localise onto
the asymptotic boundary and are classical, so they contribute no fluctuations of
their own. With standard (Dirichlet) asymptotically locally AdS boundary
conditions the leading boundary data is held fixed, and the bulk fluctuations
the shells could couple to are normalizable modes that fall off towards the
boundary; their contribution in the thin near-boundary region onto which the
shells localise is subleading in the large-mass limit, so to the order we work
the perturbative corrections reside in the central disk, where they coincide
with those of $\overline{Z(\beta)}$ around the corresponding saddle. The strip
and shell contributions cancel against the unit-norm normalisation
\eqref{eq:1sshellwh}. In this sense the bulk loop expansions around
corresponding saddles agree order by order, and the explicit trace reproduces
$\overline{Z(\beta)}$ beyond the leading saddle; the same correspondence applied
to $\overline{\left(Tr_{\mathcal{H}_{\mathcal{X}}}(e^{-\beta H})-Z(\beta)
\right)^2}$ gives $\overline{\left(Tr_{\mathcal{H}_{\mathcal{X}}}(e^{-\beta H})
-Z(\beta)\right)^2}=0$. We do not attempt to make this matching precise: it rests
on the suppression of the near-boundary fluctuations coupling to the heavy shells
in the large-mass limit, a statement we expect but do not establish here. We
therefore rely on the factorisation argument for the clean full-path-integral
equality.

\paragraph{On the large $\kappa$ limit.}
The role of the $\kappa_R\to\infty$ limit differs depending on the ensemble. In the canonical ensemble
the single-boundary Hilbert space is infinite-dimensional, so $\kappa_R\to\infty$
is genuinely required for the shell states to span; in that limit every geometry
breaking a shell index loop is suppressed, leaving no surviving $O(1/\kappa_R)$
corrections. In the micro-canonical ensemble at an energy above the black-hole
threshold the Hilbert space is finite-dimensional, of dimension
$\sim e^{S}$ with $S$ the Bekenstein-Hawking entropy, and the shell states
already span once $\kappa_R\gtrsim e^{S}$, as shown in \cite{Balasubramanian:2022gmo,Balasubramanian:2022lnw,Balasubramanian:2025zey}. For such finite $\kappa_R$ the
equality is recovered in a planar limit,
$\kappa_R,\,e^{1/G_N}\to\infty$ with $\kappa_R\,e^{-1/G_N}\sim \mathcal{O}(1)$ held fixed, in which a
 resolvent resummation (see also \cite{Penington:2019kki,Climent:2024trz}) incorporates the additional topologies; we do not carry out
this resummation explicitly, as the methods are standard and the result not
especially illuminating. In either case $Tr_{\mathcal{H}_{\mathcal{X}}}
(e^{-\beta H})=Z(\beta)$ holds because one is tracing over a genuine basis ---
shown in \cite{Balasubramanian:2025zey} for the single boundary case, to exist
micro-canonically once $\kappa_R\gtrsim e^{S}$ and canonically as
$\kappa_R\to\infty$.

\subsection{What is the Bekenstein-Hawking entropy counting?}
In thermal physics, the  micro-canonical density of states is obtained from the partition function by a Laplace transform $\rho(E) = \int d\beta \, e^{\beta E} Z(\beta)$. This density of states is often expressed as the exponential of the micro-canonical entropy. Thus, given (\ref{Tr=Z}), we should be able to read off the micro-canonical entropy in gravity in the saddlepoint approximation by  computing the Laplace transform of $\bar{Z}(\beta)$.  The path integral for $\bar{Z}(\beta)$ can have multiple saddlepoints.  The value of the entropy, and its interpretation in terms of microstates, depend on which saddlepoint dominates.  For example, in asymptotically AdS gravity $\bar{Z}(\beta)$ has three saddle geometries: the large black hole, thermal AdS and the small black hole. The small black hole is always sub-dominant, while the Hawking-Page transition determines dominance between the other two  \cite{Hawking:1982dh}. Above the transition, the Laplace transform of $\bar{Z}(\beta)$ is dominated by the black hole saddle, and the explicit saddlepoint computation yields the Bekenstein-Hawking entropy formula $S = A/4G$ where $A$ is the area of the horizon in the black hole  of energy (mass) $E$.  Similar computations lead to $S = A/4G$ in any theory which is described by Einstein gravity at low energies regardless of the cosmological constant.  This is why the Bekenstein-Hawking entropy formula is usually described as being universal.

What are the states counted by the Bekenstein-Hawking entropy formula?  Since the formula arises from a dominant saddlepoint of the Gibbons-Hawking path integral which has a black hole geometry, we expect that states contributing to the entropy can be regarded as black hole microstates.  Any such microstate should have a black hole exterior geometry, but could have structure behind the horizon that cannot be measured by the asymptotic observer. However, our derivation in Sec.~\ref{sec:facisGH} and Sec.~\ref{eq:directcal}  just required a complete basis of states, and we could have used the type B shell states, which do not contain horizons (Fig.~\ref{fig:type2}). So  what are the states counted by the Bekenstein-Hawking entropy formula, and can they be interpreted as black hole microstates?

To resolve this question, we have to examine a micro-canonical fixed energy Hilbert space $\mathcal{H}_{\mathcal{X}}|_E$.  We denote the projector into the micro-canonical Hilbert space $\mathcal{H}_{\mathcal{X}}|_E$ as $\prod_E$.  Let us denote a single-sided shell state formed by insertion of the $\mathcal{O}_i$ shell operator and time evolution by $\frac{\beta_R}{2}$  as $ \ket{i^{\beta_R}}$. Isolating the $\frac{\beta_R}{2}$ time evolution we write this as $\ket{i^{\beta_R}}=e^{-\frac{\beta_R}{2}H}\ket{\hat{i}}$. Labeling the energy eigenstates of $H$ as $\{\ket{n}\}$  we write the shell states in the energy basis as $\ket{i^{\beta_R}}= \sum_n e^{-\frac{\beta_R}{2}E_n}\braket{n|\hat{i}}\ket{n}$ and the micro-canonically projected states as $\prod_E\ket{i^{\beta_R}}= \sum_{E_n=E} e^{-\frac{\beta_R}{2}E}\braket{n|\hat{i}}\ket{n}$. States with the same $\mathcal{O}_i$  insertion but different values of $\frac{\beta_R}{2}$ therefore result in linearly dependent micro-canonical states, differing only by a factor of $e^{-\frac{\beta_R}{2}E}$. Ignoring this pre-factor, the projection can be performed by the Laplace transform  $\prod_E\ket{i}\equiv \ket{i^E} \sim \int d\beta \, e^{\beta E/2}  \ket{i^{\beta}}$. 

Any state $\ket{\Psi_E}\in \mathcal{H}_{\mathcal{X}}|_E$ can be written as a linear combination $\ket{\Psi_E}=\sum_i \alpha_i\ket{i}$ of the canonical shell states, as they span the full Hilbert space $\mathcal{H}_{\mathcal{X}}$. We must have $\prod_E\ket{\Psi_E}=\ket{\Psi_E}$ and therefore any state $\ket{\Psi_E}\in \mathcal{H}_{\mathcal{X}}|_E$ can be written as  $\ket{\Psi_E}=\sum_i \alpha_i\prod_E\ket{i}\equiv \sum_i \alpha_i\ket{i^E}$. Hence the projected shell states $\ket{i^E}$ span $\mathcal{H}_{\mathcal{X}}|_E$  because the
 canonical states $\ket{i}$ span the full Hilbert space $\mathcal{H}_{\mathcal{X}}$.

The semiclassical geometry associated to the micro-canonical shell states is that of the leading saddle contribution to its norm $\overline{\braket{i^E|i^E}}=\overline{\braket{i|\prod_E|i}}\sim \int d\beta \,  e^{\beta E} \overline{\braket{i^{\tilde{\beta}}|i^\beta}}$, where we have prepared the canonical state $\ket{i}$ at some arbitrary inverse temperature  $\frac{\tilde{\beta}}{2}$. As discussed in  Sec.~\ref{eq:directcal} there are two possible leading saddles to the canonical overlap: the Type A and Type B saddle. The Laplace transform will pick out the Type A saddle at micro-canonical energies above the black hole threshold. To see this note the canonical overlap is proportional to the Gibbons-Hawking path integral $\overline{\braket{i^{\tilde{\beta}}|i^\beta}}= Z_{m_i}\,\overline{S}(0) \,\overline{Z}(\beta + \tilde{\beta}) $ (see  Sec.~\ref{eq:directcal}). We have shown that $\overline{Z}(\beta)$ computes the thermal partition function of the gravity theory, and hence $\int d\beta \,  e^{\beta E}\, \overline{Z}(\beta + \tilde{\beta})= e^{-\tilde{\beta}E+S(E)}$ where $S(E)$ is the micro-canonical entropy. Hence $\overline{\braket{i^E|i^E}}= \overline{S}(0)  Z_{m_i}e^{-\tilde{\beta}E+S(E)}$. As shown above, for energies above the black hole threshold (if there is one) the micro-canonical entropy is given by the Bekenstein Hawking formula and therefore  the Laplace transform must have picked out the black hole saddle. Hence the geometry associated to the micro-canonical shell state is that of a Type A shell state; that is a single-boundary black hole geometry (Fig.~\ref{fig:type1}) with a shell behind the horizon, and an exterior identical to that of the black hole saddle to $\bar{Z}(\beta)$. 

Thus we see that at energies above the black hole threshold, the basis for the micro-canonical Hilbert space consists solely of black hole microstates, or, more precisely, states for which the leading saddle computing the norm is a black hole microstate.\footnote{It should  be possible to carry out a similar analysis in the context of single-sided states with End-of-the-World branes behind the horizon \cite{Geng:2024jmm}, although these have not been shown to span the full Hilbert space.} Furthermore, by similar logic the microcanonical entropy of the two-boundary theory is given by the sum of the Bekenstein-Hawking entropies of the two horizons, $S_{E_L}+S_{E_R}$, and the states that are being counted are microstates of the two-boundary black hole in the sense discussed above.  This follows from the two-boundary shell state construction in \cite{Balasubramanian:2022lnw,Balasubramanian:2022gmo,Climent:2024trz,Balasubramanian:2024rek,Balasubramanian:2024yxk} which was shown to provide a basis in \cite{Balasubramanian:2025jeu}.

\section{Entanglement entropy and traces in the gravity Hilbert space} \label{sec:ER=EPR}

Let us consider two entangled universes, namely states of a gravitating system with two disconnected Lorentzian boundaries.  Examples include the thermofield double \cite{Maldacena:2001kr} and PETS states \cite{Goel:2018ubv} in asymptotically AdS spacetime, and the two-sided black hole microstates in \cite{Balasubramanian:2022gmo,Climent:2024trz,Balasubramanian:2025jeu,Balasubramanian:2024rek}.\footnote{Entanglement can also be introduced between de Sitter universes which not have a Euclidean boundary, and asymptotically AdS or flat universes that do \cite{Balasubramanian:2021wgd, Balasubramanian:2020coy,Balasubramanian:2020xqf}, but we will not examine those cases here.}
The  full two-boundary Hilbert space $ \mathcal{H}_{\mathcal{X}_L\cup \mathcal{X}_R}$ factorises into copies of the single-boundary theories $\mathcal{H}_{\mathcal{X}_L}\otimes\mathcal{H}_{\mathcal{X}_R}$ \cite{Balasubramanian:2025zey}, and the generic state in this Hilbert space will have some nonzero entanglement between these factors. Given such a state, $|\Psi\rangle \in\mathcal{H}_{\mathcal{X}_L\cup \mathcal{X}_R}$, we  would like to compute the $n$-th Rényi entropy, 
\beq \label{eq:entropy}
\overline{Tr_{\mathcal{H}_{\mathcal{X}_R}}(\rho^n_{R})} \, ,
\eeq
where $\rho_{R}=Tr_{\mathcal{H}_{\mathcal{X}_L}}(|\Psi\rangle\langle \Psi |)$ is the reduced density matrix on $R$.\footnote{Strictly speaking the R\'{e}nyi entropy is 
$log[\overline{Tr_{\mathcal{H}_{\mathcal{X}_R}}(\rho^n_{R})}] - log[\overline{(Tr_{\mathcal{H}_{\mathcal{X}_R}} (\rho_{R}))^n}]$, 
but we will work with the simpler quantity (\ref{eq:entropy}) from which the entropy can be easily calculated, and abuse terminology to call it it R\'{e}nyi entropy. }

\begin{figure}
    \centering
    \begin{subfigure}{\linewidth}
        \centering
    \includegraphics[width=\linewidth]{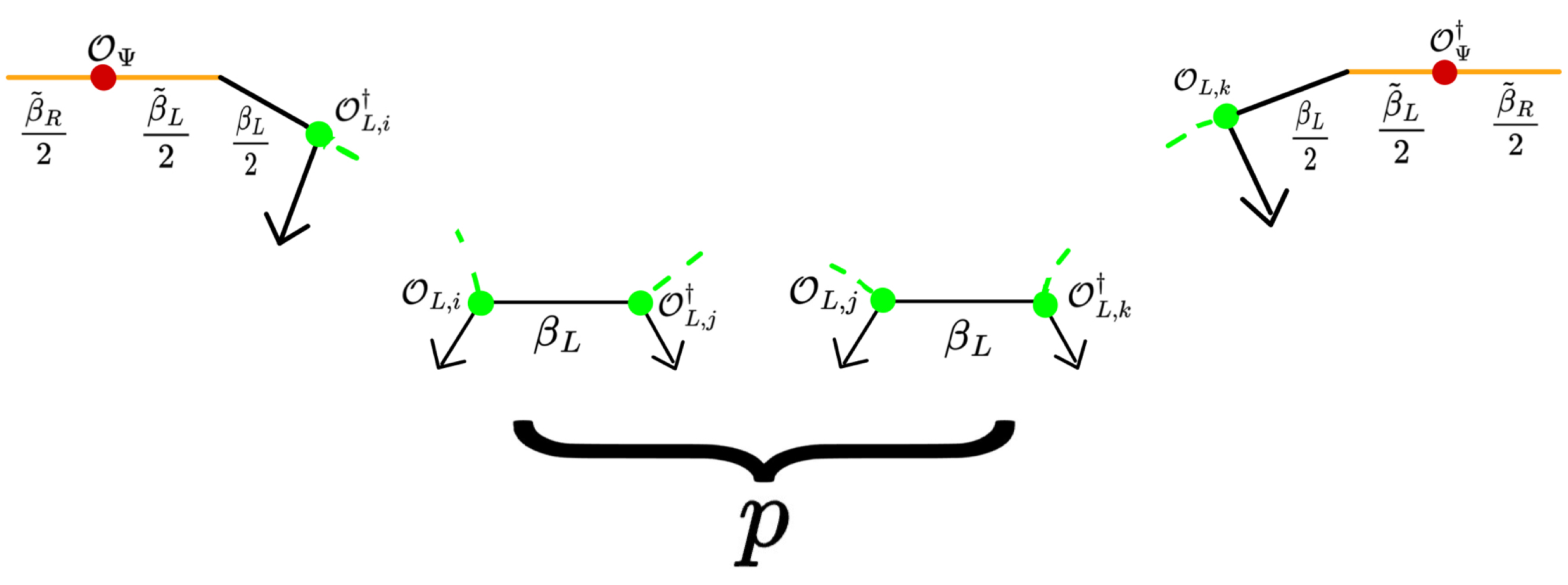}
    \caption{}
    \label{fig:rhobc}
    \end{subfigure}
\hfill
\centering
    \begin{subfigure}{\linewidth}
    \centering
    \includegraphics[width=\linewidth]{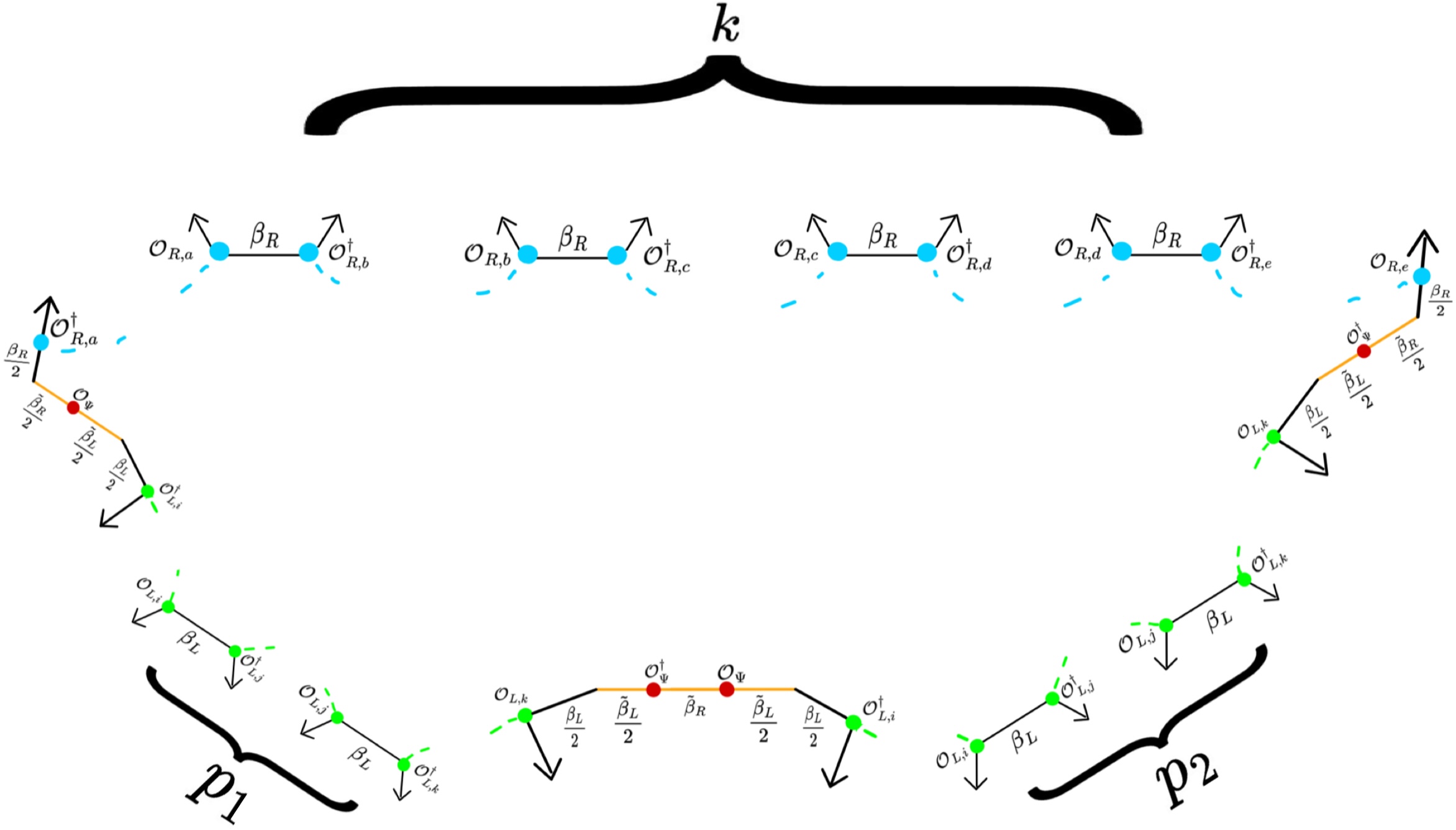}
    \caption{}
    \label{fig:traceloop}
    \end{subfigure}
    
    \caption{  ({\bf a}) Boundary condition for $\rho_{R}=Tr_{\mathcal{H}_{\mathcal{X}_L}}(|\Psi\rangle\langle \Psi |)$. ({\bf b}) Boundary condition defining $Tr_{R}(\rho_{R}^n)$, depicted for $n=2$.}
    \label{fig:Hspacerenyi1}
\end{figure}

As discussed above, $\mathcal{H}_{\mathcal{X}_{L,R}}$ are spanned by the single-sided shell states  \cite{Balasubramanian:2025zey}. We can therefore use the shell state resolution of the trace (\ref{eq:tr}) to explicitly trace out one of the universes to obtain the reduced density matrix on the other. For example, tracing out $\mathcal{H}_{\mathcal{X}_L}$ results in  $\rho_{R}=\lim_{m \to -1} (G^{m})_{L,ab} \langle b |_L\Psi\rangle\langle \Psi |a \rangle_L$ and $\rho_{R}^n=\lim_{p_{1},..p_{n} \to -1} \prod_{i=1}^{n} (G^{p_i})_{L,a_ib_i} \langle b_i |_L\Psi\rangle\langle \Psi |a_i \rangle_L$. Performing the final trace over $\mathcal{H}_{\mathcal{X}_R}$ we obtain:
\beq \label{eq:n_Rényi}
\overline{Tr_{\mathcal{H}_{\mathcal{X}_R}}(\rho^n_{R})} = \lim_{k,p_{1},..p_{n} \to -1}
\overline{(G^k)_{R,cd}\langle d|_R\prod_{i=1}^{n} (G^{p_i})_{L,a_ib_i} \langle b_i |_{L}\Psi\rangle\langle \Psi |a_i \rangle_L |c \rangle_R}
\eeq
For concreteness, and without loss of generality, we will consider states of the form $|\Psi (\tilde{\beta}_L,\tilde{\beta}_R)\rangle= |e^{-\frac{\tilde{\beta}_LH_L}{2}}\mathcal{O}_{\Psi}e^{\frac{-\tilde{\beta}_R H_R}{2}}\rangle$, where $\mathcal{O}_{\Psi}$ can be non-local and may involve time evolution.  
The boundary conditions defining the density matrix $\rho^{n}_R$ and the (unnormalised) Rényi entropy (\ref{eq:n_Rényi}) are depicted in Figs.~\ref{fig:rhobc},\ref{fig:traceloop}. We can evaluate the path integral subject to these boundary conditions explicitly within the saddlepoint approximation, and then take the ${k,p_{1},..p_{n} \to -1}$ limit to obtain the Rényi entropy.

\begin{figure}

\begin{subfigure}{\linewidth}
    \centering
    \includegraphics[width=\linewidth]{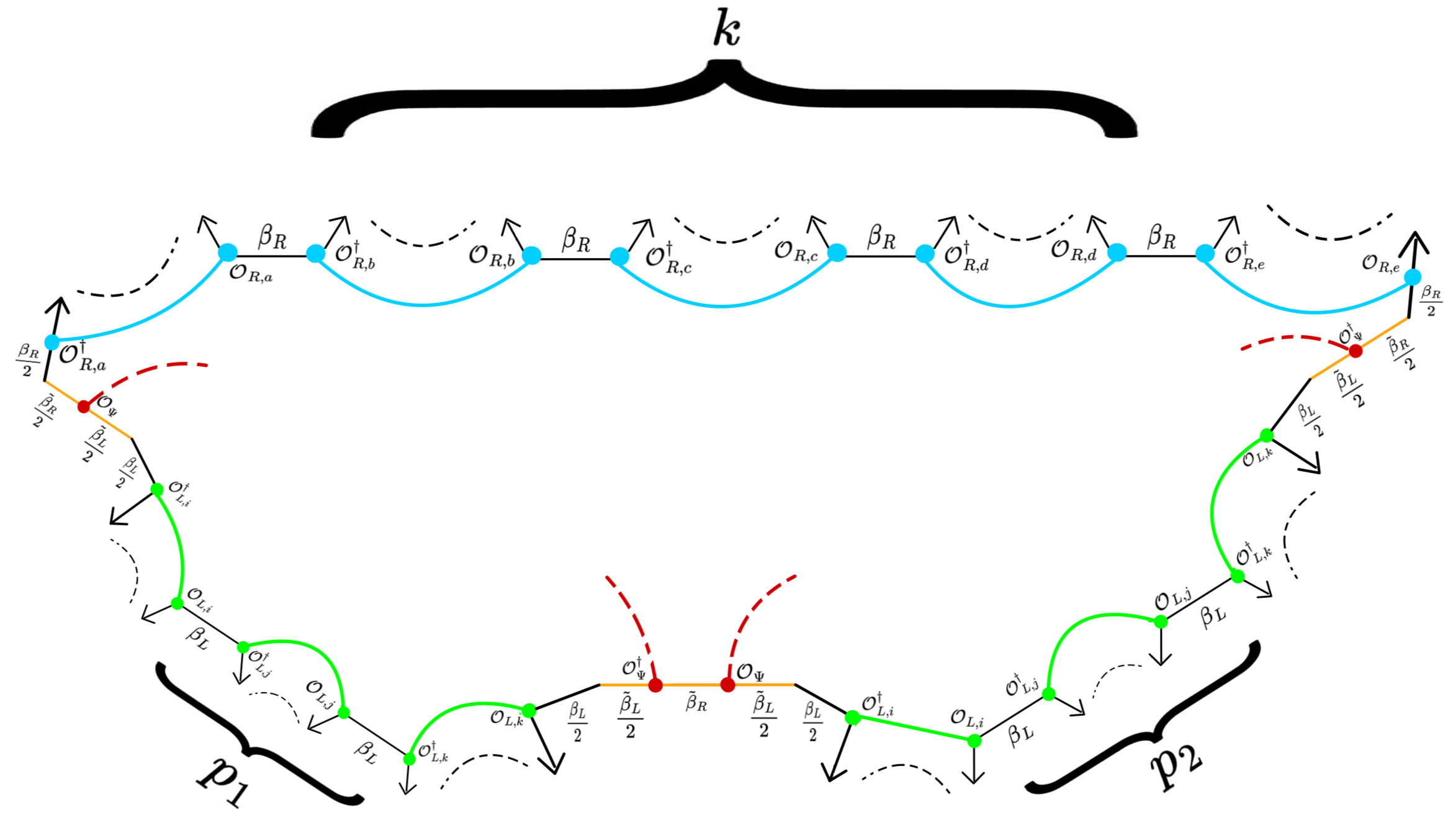}
    \caption{}
    \label{fig:entropy_disk1}
\end{subfigure}
    \begin{subfigure}{\linewidth}
    \centering
    \includegraphics[width=\linewidth]{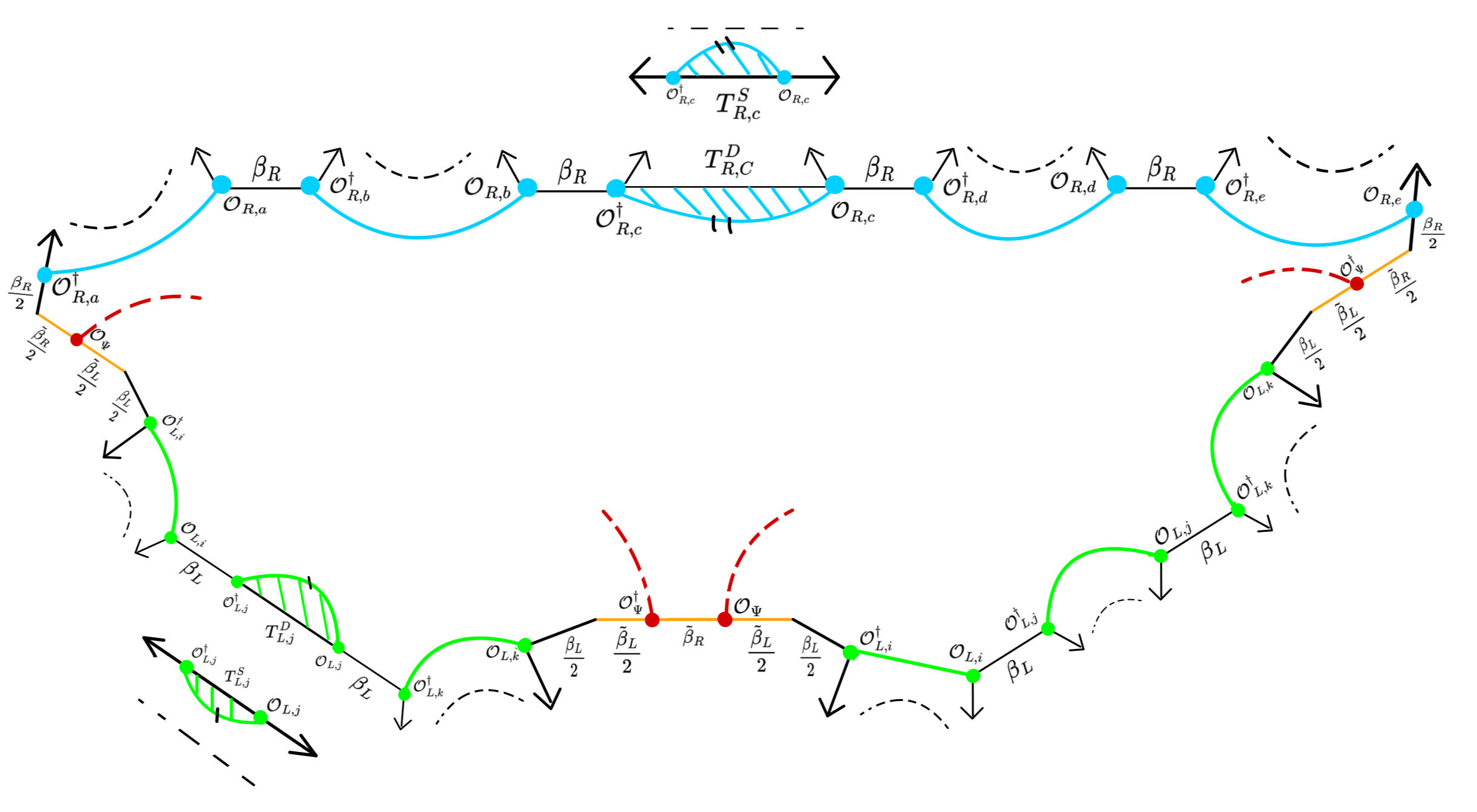}
    \caption{}
    \label{fig:entropy_disk}
\end{subfigure}
    \caption{({\bf a}) The maximally connected saddles to $\overline{(G^k)_{R,cd}\langle d|_R\prod_{i=1}^{n} (G^{p_i})_{L,a_ib_i} \langle b_i |_{L}\Psi\rangle\langle \Psi |a_i \rangle_L |c \rangle_R}$ connect all the boundaries in a single bulk.  ({\bf b}) These saddles are constructed by gluing a shell strip into the central disk along the shell-worldvolume. This is the case for all the shells, but is depicted here for one $L$ and one $R$ shell for clarity. The red dotted lines convey that this construction is agnostic about the dynamics of the $\mathcal{O}_{\Psi}$ insertions.}
    \label{fig:Hspacerenyi2}
\end{figure}

\begin{figure}
\centering
\begin{subfigure}[c]{0.45\linewidth}
    \centering
    \includegraphics[width=0.8\linewidth]{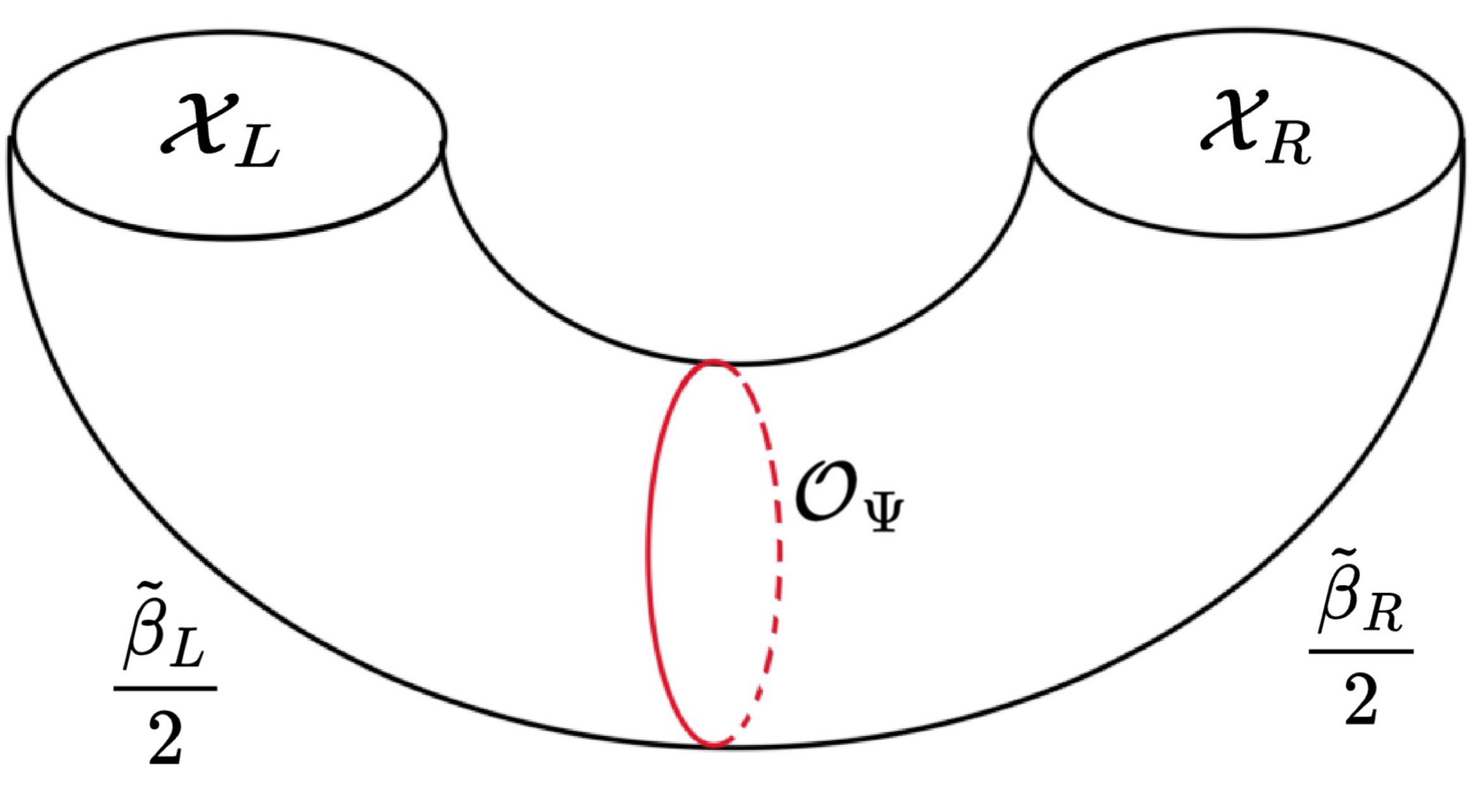}
    \caption{}
    \label{fig:1sTRBC}
\end{subfigure}
\begin{subfigure}[c]{0.45\linewidth}
    \centering
    \includegraphics[width=0.8\linewidth]{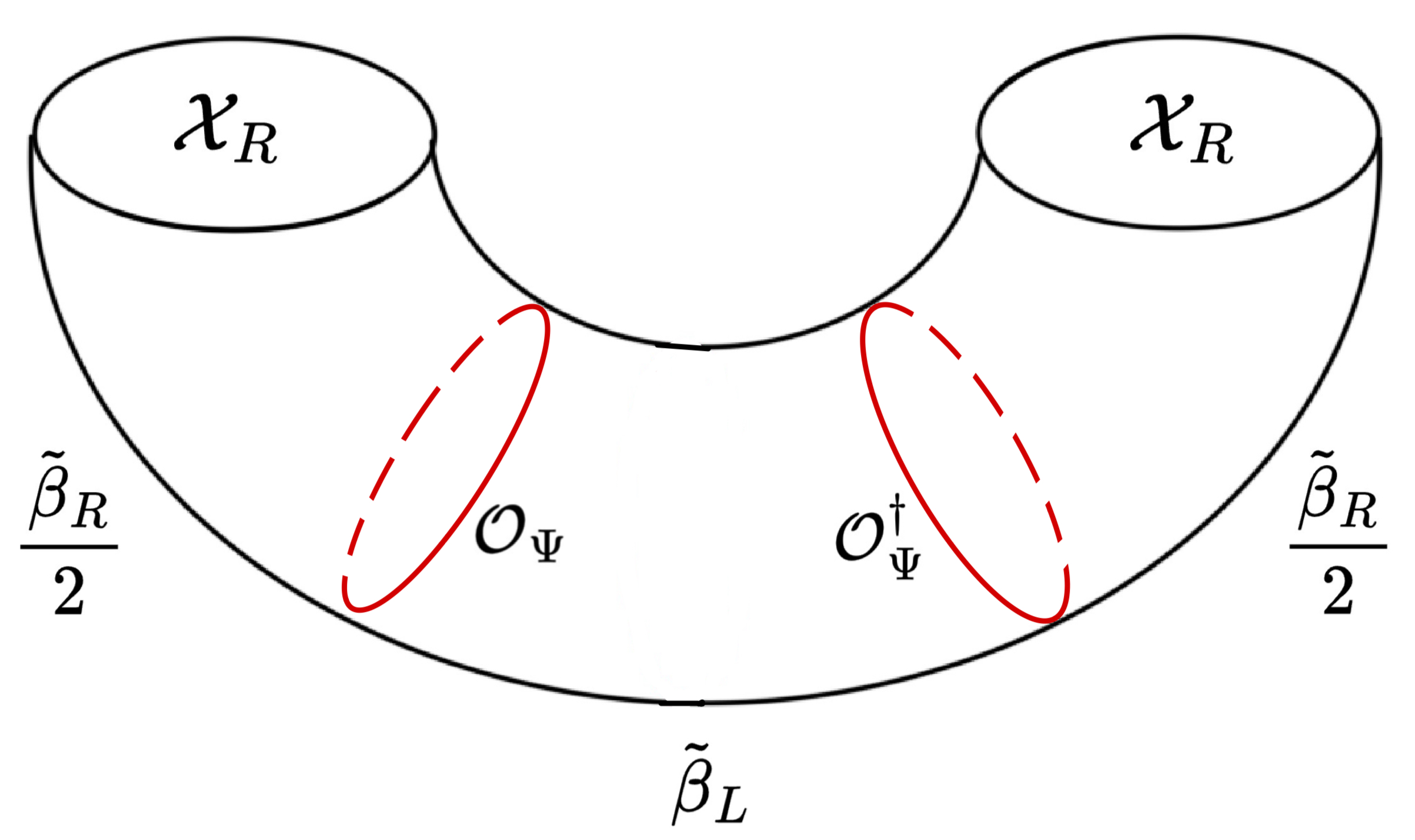}
    \caption{}
    \label{fig:denisty} 
\end{subfigure} 
    \begin{subfigure}[c]{0.45\linewidth}
    \centering
    \includegraphics[width=0.7\linewidth]{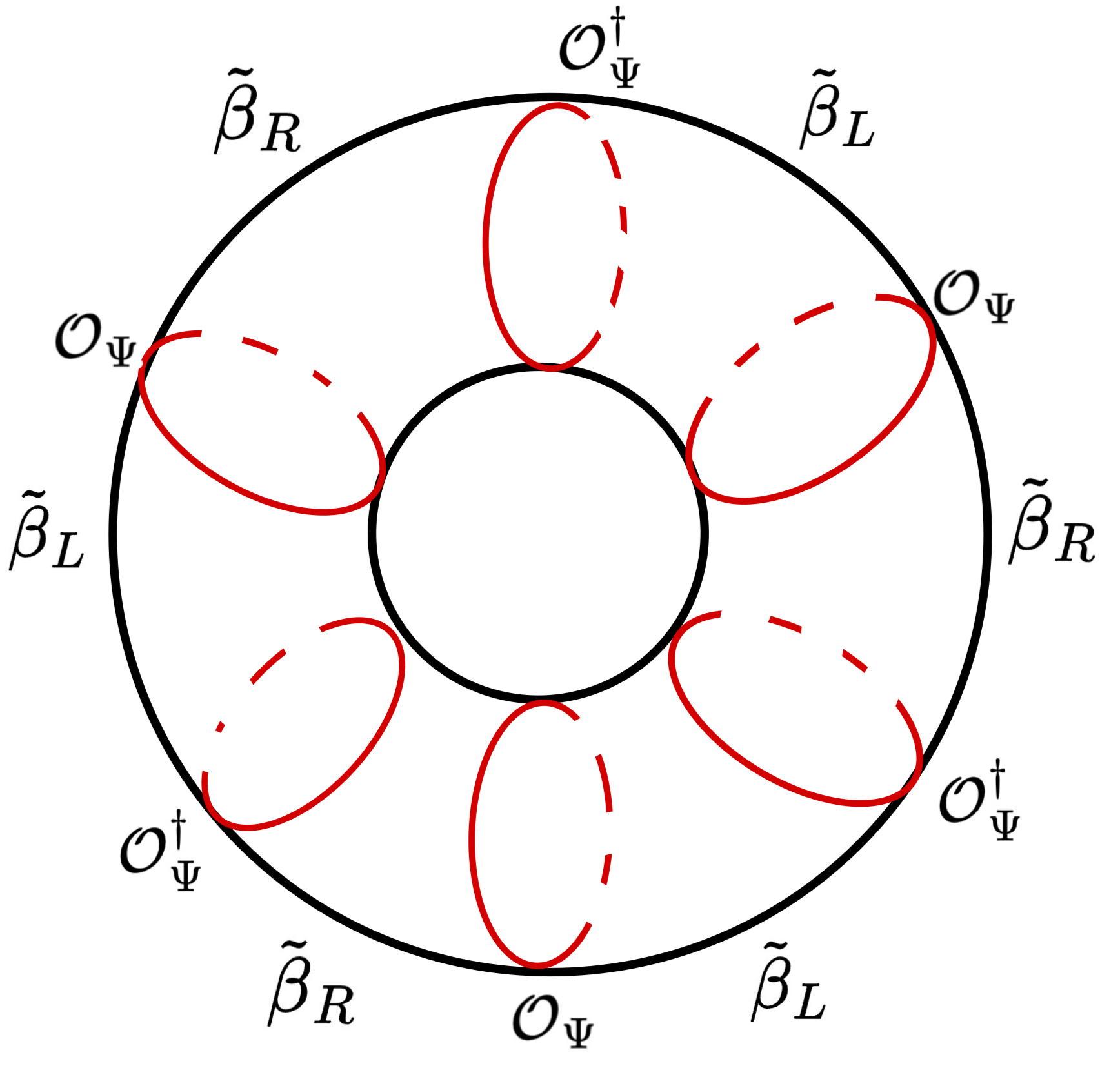}
    \caption{}
    \label{fig:TrSCWH}
    \end{subfigure}
    \caption{Path integral prescription for calculating the $n$-th Rényi entropy (\ref{eq:holotrace}) of in QFT for a state living in $\mathcal{H}_{QFT}\otimes\mathcal{H}_{QFT} $: ({\bf a}) Start with the condition defining a two-sided state $|\Psi\rangle$ which alternatively can be viewed as the operator $M_{\Psi}: \mathcal{H}_{\mathcal{X}_L} \to\mathcal{H}_{\mathcal{X}_R}$. ({\bf b}) The reduced density matrix $\rho_R=Tr_{\mathcal{H}_{\mathcal{X}_L}}(|\Psi\rangle\langle \Psi |)$ is constructed by gluing the $|\Psi\rangle$ ket to the bra $\langle \Psi |$ along the cut $\mathcal{X}_L$. This can alternatively be viewed as the operator $M_{\Psi}M_{\Psi}^{\dagger}: \mathcal{H}_{\mathcal{X}_R} \to\mathcal{H}_{\mathcal{X}_R}$. ({\bf c}) $Tr_{\mathcal{H}_{\mathcal{X}_R}}(\rho_R^n)$ is then given by gluing $n$ copies of $\rho_R$ to each-other along the $\mathcal{X}_R$ cuts and the gluing the final two $\mathcal{X}_R$ cuts together to obtain a periodic path integral. }
     \label{fig:QFTsewn}
\end{figure}

In holographic theories of gravity the boundary replica trick motivates a different boundary condition for the  path integral computing Rényi entropy. 
In particular, the holographic boundary condition for $Tr_{\mathcal{H}_{\mathcal{X}_R}}(\rho^n_{R})$, denoted  $Z_{\circ}(\rho_{\Psi}^n)(\tilde{\beta}_L,\tilde{\beta}_R)$,  corresponds to a ``torus" formed by  $n$ alternating $\mathcal{O}_{\Psi},\mathcal{O}^{\dagger}_{\Psi}$ insertions, with $\mathcal{O}_{\Psi,i},\mathcal{O}^{\dagger}_{\Psi,i}$ separated by $\tilde{\beta}_L$ and $\mathcal{O}^{\dagger}_{\Psi,i},\mathcal{O}_{\Psi,i+1}$ separated by $\tilde{\beta}_R$ boundary time (see Fig.~\ref{fig:QFTsewn}).  Indeed, in AdS/CFT this is precisely the prescription that leads to the bulk derivation \cite{Lewkowycz:2013nqa} of the RT and HRT formulae \cite{Ryu:2006bv,Ryu:2006ef,Hubeny:2007xt}  relating the entanglement entropy of  a boundary subregion in the CFT to area of an extremal surface in the bulk homologous to that subregion.  This leads to a puzzle analogous to one we discussed in Sec.~\ref{sec:TPF} for the Gibbons-Hawking path integral:  
{\it Why does the holographic replica path integral prescription for the R\'{e}nyi entropy (Fig.~\ref{fig:TrSCWH}) reproduce the trace over the Hilbert space computed by the sum over path integrals with boundary conditions  in Fig.~\ref{fig:Hspacerenyi1}?
}

Concretely,  we must show that the gravity Hilbert space prescription and the holographic boundary replica prescription for computing the R\'{e}nyi entropy are the same at the fine-grained level:
\beq \label{eq:holotrace}
Tr_{\mathcal{H}_{\mathcal{X}_R}}(\rho^n_{R}) \stackrel{?}{=}Z_{\circ}(\rho_{\Psi}^n)(\tilde{\beta}_L,\tilde{\beta}_R) \, .
\eeq
Comparing Fig.~\ref{fig:Hspacerenyi1} and Fig.~\ref{fig:Hspacerenyi2}  with Fig.~\ref{fig:QFTsewn} we want to show from the gravity Hilbert space perspective that: ({\bf 1}) the trace over $\mathcal{H}_{\mathcal{X}_L}$ in the definition of the reduced density matrix $\rho_R$ sews the $\Psi$ bra and ket together along the $\mathcal{X}_L$ cut to produce a  ``cylinder," and ({\bf 2}) the trace over $\mathcal{H}_{\mathcal{X}_R}$  in $ Tr_{\mathcal{H}_{\mathcal{X}_R}}(\rho^n_{R})$ sews  copies  of the  cylinder  together into the ``torus" of Fig.~\ref{fig:TrSCWH}. Analogously to the discussion in Sec.~\ref{sec:puzzles} these two requirements are not immediate from the rules for cutting the gravitational path integral to obtain states that were outlined at the beginning of Sec.~\ref{sec:QTFTPI}.

We will  provide two arguments showing (\ref{eq:holotrace}). The first uses strategic insertions of the identity to show ({\bf 1}) is true in general, and show requirement ({\bf 2}) from factorisation of the two-boundary Hilbert space. Second, we  evaluate the traces in (\ref{eq:n_Rényi}) explicitly in the single-sided shell basis and find equality between the two prescriptions.

 \subsection{The trace glues density matrices}
 Consider the boundary condition defining the reduced density matrix (Fig.~\ref{fig:rhobc})
 \beq \label{eq:reduce}
 \rho_{R}=Tr_{\mathcal{H}_{\mathcal{X}_L}}(|\Psi\rangle\langle \Psi |)=\lim_{m \to -1} (G^{m})_{L,ab} \langle b |_L\Psi\rangle\langle \Psi |a \rangle_L \, .
 \eeq
 We will argue that the role of the trace over the left Hilbert space is to sew the the $L$ parts of the $\ket{\Psi}$ bra and ket together into a boundary condition given by $\frac{\tilde{\beta}_R}{2}$ time evolution, $O_{\Psi}$ insertion,  $\tilde{\beta}_L$ time evolution, $O_{\Psi}^{\dagger}$ insertion and finally another $\frac{\tilde{\beta}_R}{2}$ time evolution (Fig.~\ref{fig:denisty}).  To see this, note that the state $|\Psi\rangle$ can alternatively be considered as a map $M_{\Psi}: \mathcal{H}_{\mathcal{X}_L} \to\mathcal{H}_{\mathcal{X}_R}$ that starts with a  state in $\mathcal{H}_{\mathcal{X}_L}$, time evolves by $\frac{\tilde{\beta}_L}{2}$, applies to operator $ O_{\Psi}$ and then time evolves by $\frac{\tilde{\beta}_R}{2}$ up to the cut $\mathcal{X}_R$ to define a state in $\mathcal{H}_{\mathcal{X}_R}$. Similarly, $M_{\Psi}^{\dagger}$ takes in a state in $\mathcal{H}_{\mathcal{X}_R}$, time evolves by $\frac{\tilde{\beta}_R}{2}$, applies $ O^{\dagger}_{\Psi}$ and finally time evolves by $\frac{\tilde{\beta}_L}{2}$ to define a state in $\mathcal{H}_{\mathcal{X}_L}$. In this notation the above ``glued" boundary condition is produced by $M_{\Psi}M_{\Psi}^{\dagger}$.

 To make progress let us insert the identity to write
 $M_{\Psi}M_{\Psi}^{\dagger}=M_{\Psi}\mathds{1}_{\mathcal{H}_{\mathcal{X}_L}}M_{\Psi}^{\dagger}$. We resolve this insertion of the identity in a basis ``conjugate" to the one used for the trace over $\mathcal{H}_{\mathcal{X}_L}$, defined as follows. Given a basis set $\{\ket{a}_L\}$ defined by operator insertions $O_{a}$ on some cut manifold $\mathcal{M}_a$ with cut $\mathcal{X}$, the conjugate basis $\{\ket{\tilde{a}}_L\}$ is defined by placing the operator $O_{a}^{\dagger}$ in place of $O_{a}$, hence $\ket{\tilde{a}} \equiv \ket{a^*}$. Note that if $\{\ket{a}_L\}$  forms a basis then so must $\{\ket{\tilde{a}}_L\}$. Resolving the identity in this conjugate basis we obtain:
 \beq \label{eq:gluedbraket}
 M_{\Psi}\mathds{1}_{\mathcal{H}_{\mathcal{X}_L}}M_{\Psi}^{\dagger}= \lim_{m\to -1} (G^m)_{L,b^*a^*}M_{\Psi} \ket{b^*}_L\bra{a^*}_LM_{\Psi}^{\dagger} \, .
\eeq

Now, for any two states $\ket{m}_R,\ket{n}_R \in\mathcal{H}_{\mathcal{X}_R}$ we have $\bra{m}_R M_{\Psi} \ket{b^*}_L= \bra{m}_R \bra{b}_L\ket{\Psi}$ and similarly $\bra{a^*}_LM_{\Psi}^{\dagger}\ket{n}_R=\bra{\Psi}\ket{a}_L\ket{n}_R$. By definition we also have $G_{b^*a^*}=\braket{b^*|a^*}=\braket{a|b}=G_{ab}$ and therefore the matrix elements of (\ref{eq:reduce}) and (\ref{eq:gluedbraket}) are equal. It follows that $\rho_{R} =M_{\Psi}M_{\Psi}^{\dagger}$ and hence the reduced density matrix on $\mathcal{H}_{\mathcal{X}_R}$ is indeed obtained by sewing the ket and bra together, showing requirement ({\bf 1}). This allows us to use $n$-fold concatenation of the sewn boundary condition $\rho_{R}^n=(M_{\Psi}M_{\Psi}^{\dagger})^n$ instead of the more involved boundary condition $\rho_{R}^n=\lim_{m_{1},..m_{n} \to -1} \prod_{i=1}^{n} (G^{m_i})_{L,a_ib_i} \langle b_i |_L\Psi\rangle\langle \Psi |a_i \rangle_L$ to compute Rényi entropies.

\paragraph{Factorisation sews the trace.}
Having shown that the trace over $\mathcal{H}_{\mathcal{X}_L}$ sews the bra and ket together into $M_{\Psi}M_{\Psi}^{\dagger}$ we can compute the trace as $Tr_{\mathcal{H}_{\mathcal{X}_R}}(\rho^n_{R})= Tr_{\mathcal{H}_{\mathcal{X}_R}}(\prod_{i=1}^{n}M_{\Psi,i}M_{\Psi,i}^{\dagger})$. We can rewrite this by inserting the identity on $\mathcal{H}_{\mathcal{X}_L}$ in between the $k$-th $M_{\Psi}M_{\Psi}^{\dagger}$ insertion, resulting in $ Tr_{\mathcal{H}_{\mathcal{X}_R}}(\rho^n_{R})= Tr_{\mathcal{H}_{\mathcal{X}_R}}(M_{\Psi,1}M_{\Psi,1}^{\dagger}\cdots M_{\Psi,k}\mathds{1}_{\mathcal{H}_{\mathcal{X}_L}}M_{\Psi,k}^{\dagger} \cdots  M_{\Psi,n}M_{\Psi,n}^{\dagger})$. Resolving the trace in the $\mathcal{H}_{\mathcal{X}_R}$ conjugate basis we obtain
 \beq
 Tr_{\mathcal{H}_{\mathcal{X}_R}}(\rho^n_{R})=  
\lim_{x,y \to -1} (G^x)_{R,a^*b^*}(G^y)_{L,st} \bra{b^*}_RM_{\Psi,1}M_{\Psi,1}^{\dagger}\cdots M_{\Psi,k}\ket{s}_L\bra{t}_LM_{\Psi,k}^{\dagger} \cdots  M_{\Psi,n}M_{\Psi,n}^{\dagger}\ket{a^*}_R \, .
\eeq

\begin{figure}
    \centering
    \includegraphics[width=1\linewidth]{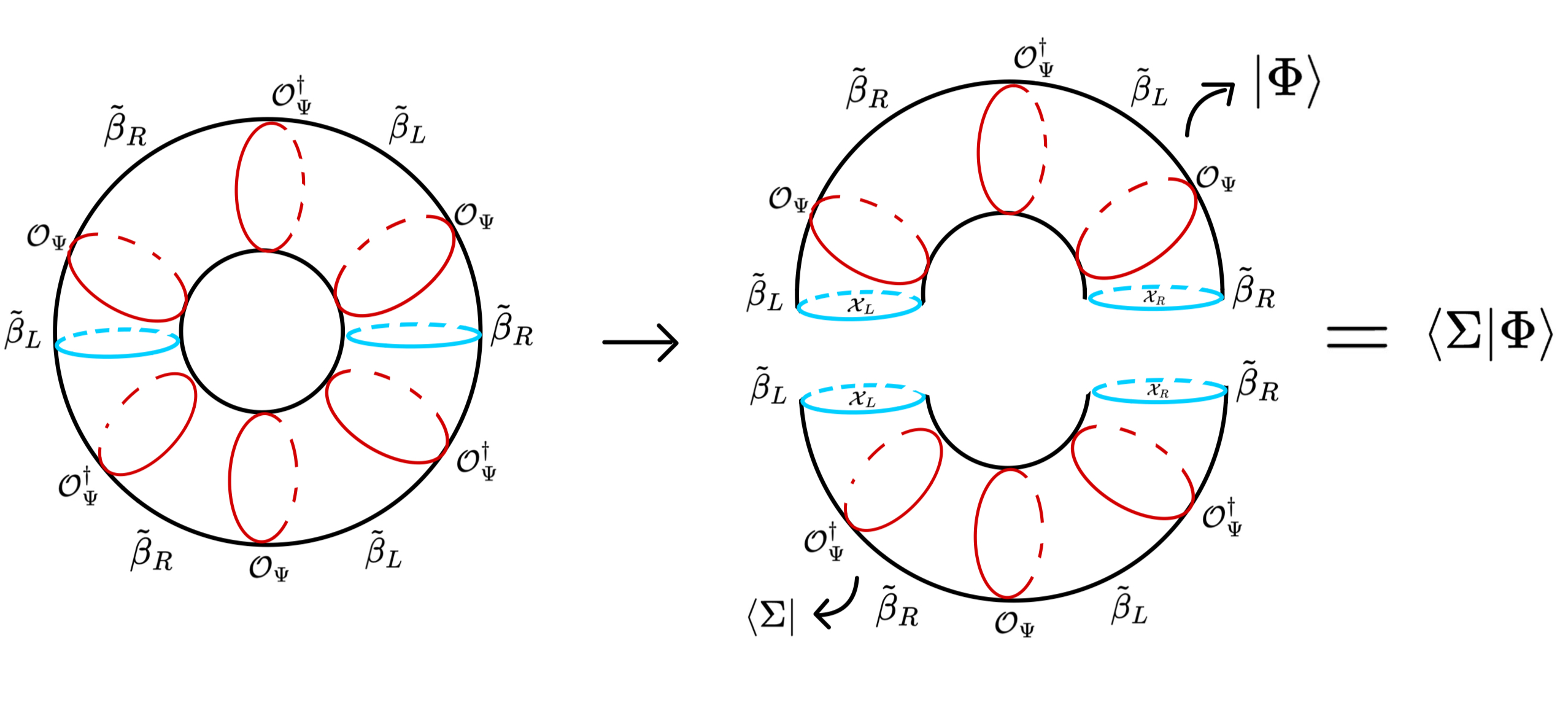}
    \caption{The path integral $\overline{Z_{\circ}(\rho_{\Psi}^n)(\tilde{\beta}_L,\tilde{\beta}_R)}$ can be re-written as the overlap $\overline{\braket{\Sigma|\Phi}}$ by cutting it into two parts along cuts $\mathcal{X}_L$ and $\mathcal{X}_R$.}
    \label{fig:renyioverlap}
\end{figure}

Note that $M_{\Psi,1}M_{\Psi,1}^{\dagger}\cdots M_{\Psi,k}$ and $M_{\Psi,k}^{\dagger} \cdots  M_{\Psi,n}M_{\Psi,n}^{\dagger}$ can be seen as a bra $\bra{\Sigma}$ and ket $\ket{\Phi}$  on $\mathcal{H}_{\mathcal{X}_L \cup\mathcal{X}_R} $ respectively. In this notation the path integral with the boundary condition defining $Z_{\circ}(\rho_{\Psi}^n)(\tilde{\beta}_L,\tilde{\beta}_R)$ computes the overlap between these states (Fig.~\ref{fig:renyioverlap})
\beq \label{eq:5}
Z_{\circ}(\rho_{\Psi}^n)(\tilde{\beta}_L,\tilde{\beta}_R)= \braket{\Sigma|\Phi} \, .
\eeq 

Furthermore we have 
$\bra{b^*}_R M_{\Psi,1}M_{\Psi,1}^{\dagger}\cdots M_{\Psi,k}\ket{s}_L= \bra{\Sigma}\ket{s}_L\ket{b}_R$ and  $\bra{t}_L M_{\Psi,k}^{\dagger} \cdots  M_{\Psi,n}M_{\Psi,n}^{\dagger}\ket{a^*}_R=\bra{t}_L \bra{a}_R\ket{\Phi}$
and $G_{R,a^*b^*}=G_{R,ba}$. Using this we obtain:
\beq\label{eq:1}
 Tr_{\mathcal{H}_{\mathcal{X}_R}}(\rho^n_{R})= \lim_{x,y \to -1}G_{R,ba}(G^y)_{L,st} \bra{\Sigma}\ket{s}_L\ket{b}_R\bra{t}_L \bra{a}_R\ket{\Phi}=  \braket{\Sigma| \mathds{1}_{\mathcal{X}_L}\otimes \mathds{1}_{\mathcal{X}_R}|\Phi}= \braket{\Sigma|\Phi}
\eeq
where  the last equality follows from the factorisation (\ref{eq:facprob}) of the two-boundary Hilbert space $\mathcal{H}_{\mathcal{X}_L \cup\mathcal{X}_R} $. As in Sec.~\ref{sec:facisGH}, once factorisation~(\ref{eq:facprob}) is granted this last step is a consequence of linear algebra: the gravitational content resides entirely in the factorisation result of \cite{Balasubramanian:2025zey}, and this subsection should not be read as an independent gravitational derivation. Hence the two distinct  boundary conditions (\ref{eq:n_Rényi}) and $Z_{\circ}(\rho_{\Psi}^n)(\tilde{\beta}_L,\tilde{\beta}_R)$ for the path integral computing Rényi entropy  define the same gravitational path integral and (\ref{eq:holotrace}) holds in the fine-grained theory. Here we have shown requirement ({\bf 2});  i.e., for any operator $\mathcal{U}:\mathcal{H}_{\mathcal{X}_R} \to  \mathcal{H}_{\mathcal{X}_R}$ the role of the trace $Tr_{\mathcal{H}_{\mathcal{X}_R}}(\mathcal{U})$ is to sew the boundary condition defining the operator together along the two $\mathcal{X}_R$ cuts. 
This sewing is exactly what we expect if there were a quantum theory living on the boundary that is dual to the bulk gravitational system. Interestingly, the derivation given here relies only on the factorisation of the two boundary Hilbert space, which was shown in \cite{Balasubramanian:2025zey} to hold in asymptotically flat quantum gravity too.

\subsection{Explicitly tracing out a universe}
We now use the single-sided shell basis (see Sec.~\ref{eq:directcal}) to explicitly evaluate the traces in (\ref{eq:n_Rényi}).  In particular, we apply the toolkit of \cite{Balasubramanian:2025jeu} to show that $\overline{\mathcal{A} - \mathcal{B}}{}=0$ and $\overline{(\mathcal{A} - \mathcal{B})^2}{}=0$ with
$\mathcal{A}=Tr_{\mathcal{H}_{\mathcal{X}_R}}(\rho^n_{R})$ and $\mathcal{B}=Z_{\circ}(\rho_{\Psi}^n)(\tilde{\beta}_L,\tilde{\beta}_R)$; together these give the fine-grained equality $\mathcal{A}=\mathcal{B}$, that is~(\ref{eq:holotrace}). The boundary condition for $\rho_{R}^n$ consists of concatenating $n$ copies of the $\rho_{R}$  boundary condition as in Fig.~\ref{fig:rhobc}. The trace over $\mathcal{H}_R$ consists of adjoining an $R$ shell state boundary to each of the two ends of $\rho_{R}^n$, thereby forming a ``complete" boundary condition. These $R$ shell insertions form an index loop with the $k$ $R$-shell boundary insertions from the $(G^k)_R$ insertion, see Fig.~\ref{fig:traceloop}.

\paragraph{The averaged equality.}
We consider the $\kappa_L,\kappa_R \to \infty$ limit in which the single-sided shell states span. As explained in \cite{Balasubramanian:2025jeu} in these limits only  geometries that do not break any  index loops in (\ref{eq:n_Rényi}) contribute. The mechanism is the same as for the thermal trace in Sec.~\ref{eq:directcal}: a geometry that breaks an index loop carries fewer free shell-flavour sums than one that preserves it, and is therefore suppressed by powers of $1/\kappa$ in the spanning limit. The surviving geometries are those in which each shell worldline is contracted with its conjugate, so that the $L$- and $R$-shell index loops of~(\ref{eq:n_Rényi}) are realised in the bulk by a single connected geometry. In the relevant saddle points for (\ref{eq:n_Rényi}) at a given $\{k,p_i\}$, the shell index loops are reflected in the  bulk by a loop of boundary and shell world volume segments (Fig.~\ref{fig:entropy_disk1}).  These saddle points to (\ref{eq:n_Rényi}) are  constructed analogously to the wormhole saddles  in Sec.~\ref{eq:directcal}; that is, we glue the interior portion of a disk with shell and $\mathcal{O}_{\Psi},\mathcal{O}^{\dagger}_{\Psi}$ operator insertions on the asymptotic boundary
to $ k+ \sum_{i=1}^n p_i$ shell strips. Each shell strip carries a single shell propagating between an $\mathcal{O}_S$ and an $\mathcal{O}^{\dagger}_S$ insertion; gluing it to the disk along that worldvolume and discarding the shell homology region on either side (Fig.~\ref{fig:entropy_disk}) contracts one shell index loop, so that the $k+\sum_i p_i$ strips realise all the index contractions of~(\ref{eq:n_Rényi}). The $L$ shells propagate through for a boundary time $T_{L,i}$ on the disk boundary and a time $T^{S}_{L,i}$ on the strip boundary, and similarly for the $R$ shells. The Israel junction conditions glue these worldvolumes and fix the propagation times $\Delta T_{L_i,j},\Delta T_{R,j}$ and their strip counterparts, exactly as for the wormhole saddles of Sec.~\ref{eq:directcal}. Apart from the shell insertions, the operator insertions on the boundary of the disk are similar to that of $Z_{\circ}(\rho_{\Psi}^n)(\tilde{\beta}_L,\tilde{\beta}_R)$, where instead the $i\neq n$-th $\mathcal{O}_{\Psi},\mathcal{O}^{\dagger}_{\Psi}$ operator pair is separated by a boundary time $\tilde{\beta}_L+(p_i+1) \beta_L + \sum_{j=1}^{p_i+1}\Delta T_{L_i, j}$ and  $\mathcal{O}_{\Psi,1},\mathcal{O}^{\dagger}_{\Psi,n}$ separated by $\tilde{\beta}_R+(k+1)\beta_R + \sum_{j=1}^{k+1}\Delta T_{R, j}$.

In the large shell mass limit, all the shell propagation times $\Delta T_i,\Delta T^S_i \to 0$ go to zero, the shell homology regions on the disk and strips pinch off and the shells contribute universally to the action. The contribution from all the $L$ shells in this limit is $\prod_{i=1}^{n}\prod_{j=1}^{p_i+1} Z_{m_j,L}$ and that for the $R$ shells $\prod_{l=1}^{k+1}Z_{m_l,R}$. As $\Delta T^S_i \to 0$ the contribution from each strip is simply that of the strip of length zero $\overline{S}(0)$, resulting in a factor of $\overline{S}(0)^{k+1}\prod_{i=1}^{n}\overline{S}(0)^{p_i+1}$.  These shell and strip terms all cancel after normalizing the shell states to unit norm (see Eq.~\ref{eq:1sshellwh}). Explicitly, each normalised shell carries a factor $1/\sqrt{Z_{m}\,\overline{Z}(\beta)\,\overline{S}(0)}$ from~(\ref{eq:1sshellwh}), and there are precisely as many such factors as there are shell and strip insertions, so the product of $Z_{m}$ and $\overline{S}(0)$ factors is cancelled. Lastly, the contribution to the action from the central disk section in this limit is that of the saddle geometries satisfying the periodic asymptotic boundary condition where the $i\neq n$-th $\mathcal{O}_{\Psi},\mathcal{O}^{\dagger}_{\Psi}$ operator pair is separated by a boundary time $\tilde{\beta}_L+(p_i+1) \beta_L$ and $\mathcal{O}_{\Psi,1},\mathcal{O}^{\dagger}_{\Psi,n}$ separated by $\tilde{\beta}_R+(k+1)\beta_R$. In the ${k,p_{1},\cdots p_{n} \to -1}$ limit this disk boundary condition becomes precisely $Z_{\circ}(\rho_{\Psi}^n)(\tilde{\beta}_L,\tilde{\beta}_R)$ (see Fig.~\ref{fig:entropy_disk}), so that $\overline{\mathcal{A}}=\overline{\mathcal{B}}$.

\paragraph{The variance and cross-replica wormholes.}
To upgrade the equality of averages to the fine-grained equality~(\ref{eq:holotrace}) we must also show $\overline{(\mathcal{A}-\mathcal{B})^2}=0$. Expanding the square,
\beq \label{eq:Rvariance}
\overline{(\mathcal{A}-\mathcal{B})^2}=\overline{\mathcal{A}^2}-2\,\overline{\mathcal{A}\,\mathcal{B}}+\overline{\mathcal{B}^2}\, ,
\eeq
so it suffices to show that $\overline{\mathcal{A}^2}$, $\overline{\mathcal{A}\,\mathcal{B}}$ and $\overline{\mathcal{B}^2}$ all agree. Each is a sum of a disconnected piece --- the product of the two means, which already agree by the result above --- and connected pieces in which bulk geometries link the two replicas. It is therefore enough to match these connected, wormhole contributions across the three quantities.

Consider $\overline{\mathcal{A}\times\mathcal{B}}$. Whether connected saddles linking the two replicas exist depends on the operator $\mathcal{O}_{\Psi}$ defining the state. If the  quantity $\overline{\mathcal{B}\times\mathcal{B}}=\overline{Z_\circ\times Z_\circ}$ has no connected saddle, then neither does $\overline{\mathcal{A}\times\mathcal{B}}$, and the three terms in~(\ref{eq:Rvariance}) are trivially equal. When connected saddles do exist, they are built exactly as the single-replica saddles above were: by gluing the shell strips of the $\mathcal{A}$ replica into the fully-connected saddles of $\overline{Z_\circ\times Z_\circ}$ along the shell worldvolumes (this is the analogue of the construction in \cite{Balasubramanian:2025jeu,Balasubramanian:2025zey}; see for example Fig.~18 in \cite{Balasubramanian:2025zey}). The strips again shift the boundary times of the $\mathcal{A}$ replica's $Z_\circ$ factor by the shell preparation temperatures and propagation times. In particular, if there exist fully connected contributions to $\overline{ Z_{\circ}(\rho_{\Psi}^n)(\tilde{\beta}_L,\tilde{\beta}_R) \times Z_{\circ}(\rho_{\Psi}^n)(\tilde{\beta}_L,\tilde{\beta}_R)}$ then the additional wormhole saddles relevant for $\overline{Tr_{\mathcal{H}_{\mathcal{X}_R}}(\rho^n_{R})\times  Z_{\circ}(\rho_{\Psi}^n)(\tilde{\beta}_L,\tilde{\beta}_R)}$ are constructed by gluing shell strips into the fully-connected saddles to 
\beq \label{eq6}
\overline{Z_{\circ}(\rho_{\Psi}^n)(\tilde{\beta}_L+(p_i+1) \beta_L + \sum_{j=1}^{p_i+1}\Delta T_{L_i, j},\tilde{\beta}_R+(k+1)\beta_R + \sum_{j=1}^{k+1}\Delta T_{R, j})\times Z_{\circ}(\rho_{\Psi}^n)(\tilde{\beta}_L,\tilde{\beta}_R)} \, . 
\eeq 
This time evolved $Z_{\circ}$ boundary condition accommodates the propagation times of the shells, which go to zero as $m_i \to \infty$, and the $L,R$ shell inverse preparation temperatures ${\beta}_L,{\beta}_R$. Taking the large shell-mass limit exactly as for the mean sends the propagation times to zero and cancels the strip and shell factors against the unit-norm normalisation, so the shifted arguments of the first $Z_\circ$ in~(\ref{eq6}) reduce to $\tilde{\beta}_L,\tilde{\beta}_R$. In the ${k,p_{1},\cdots p_{n} \to -1}$ limit each of the fully connected saddles to (\ref{eq6}) limits to the connected contribution to  $\overline{ Z_{\circ}(\rho_{\Psi}^n)(\tilde{\beta}_L,\tilde{\beta}_R) \times Z_{\circ}(\rho_{\Psi}^n)(\tilde{\beta}_L,\tilde{\beta}_R)}$$\,=\overline{\mathcal{B}^2}$. The identical construction with the second replica also prepared as a shell-trace produces the connected saddles of $\overline{\mathcal{A}^2}$, which limit to the same object. Hence $\overline{\mathcal{A}^2}=\overline{\mathcal{A}\,\mathcal{B}}=\overline{\mathcal{B}^2}$, and $\overline{(\mathcal{A}-\mathcal{B})^2}=0$. This shows  $Tr_{\mathcal{H}_{\mathcal{X}_R}}(\rho^n_{R})= Z_{\circ}(\rho_{\Psi}^n)(\tilde{\beta}_L,\tilde{\beta}_R)$ in the fine-grained theory, at least within the saddlepoint approximation. 

We may again extend this analysis beyond the leading saddle. As with the
thermal trace, the cleanest statement is the factorisation argument outlined
above: we showed there that $Tr_{\mathcal{H}_{\mathcal{X}_R}}(\rho^n_R)$ and
$\langle\Sigma|\mathds{1}_{\mathcal{X}_L}\otimes\mathds{1}_{\mathcal{X}_R}|\Phi
\rangle$ represent path integrals with the same boundary conditions (see the
discussion around~\eqref{eq:5}), and that, once the two-boundary Hilbert space
factorises, $Tr_{\mathcal{H}_{\mathcal{X}_R}}(\rho^n_R)=\langle\Sigma|\Phi\rangle
=Z_{\circ}(\rho_{\Psi}^n)(\tilde{\beta}_L,\tilde{\beta}_R)$ follows purely
algebraically; this is therefore valid for the full path integral whenever it is
suitably defined, independently of any saddle expansion. The explicit shell-basis
evaluation above provides an independent check, which can be extended beyond the
leading saddle by the same heuristic argument as in Sec.~\ref{eq:directcal}: the
saddles of the two sides are in one-to-one correspondence, and in the large
shell-mass limit the strips decouple while the heavy shells localise near the
asymptotic boundary, where with standard boundary conditions the fluctuations
they couple to are normalizable and subleading, so that to the order we work the
perturbative corrections reside in the central region and match those of
$Z_{\circ}(\rho_{\Psi}^n)$ order by order. As before, we do not make this second
argument precise --- it rests on the suppression of the near-boundary
fluctuations in the large-mass limit --- and we rely on the factorisation
argument for the clean full-path-integral statement.

\section{Summary and discussion} \label{sec:Summary}

In this paper we have explained why the Euclidean path integral with one asymptotic boundary and periodic time computes a thermal trace over the gravity Hilbert space, as Gibbons and Hawking proposed. 
We have shown equality between these \textit{a priori} different gravitational path integrals in two ways. First we cut the Gibbons-Hawking path integral in half to write it as the overlap of a two-boundary gravity state and inserted a resolution of the two-boundary identity in between the bra and ket. We showed that this expression is equal to the single-boundary thermal trace if the two-boundary Hilbert space factorises into copies of the single-boundary theories, which was established in \cite{Balasubramanian:2025zey}. This argument is completely general and did not rely on a particular form of the gravitational action.  Our second argument used a particular basis for the single-boundary Hilbert space to compute the thermal trace explicitly, and again found equality between the two prescriptions. Interestingly, in either method nonpertubative gravitational effects are crucial, suggesting the Gibbons-Hawking path integral does not equal the thermal trace in  perturbative gravity. 

While the expression $Tr_{\mathcal{H}_{\mathcal{X}}}(e^{-\beta H})=\langle \beta|\mathds{1}_{\mathcal{X}_L} \otimes \mathds{1}_{\mathcal{X}_R}|\beta\rangle $
derived around (\ref{equality}) shows that the factorisation $\mathcal{H}_{\mathcal{X}_L \cup\mathcal{X}_R}=\mathcal{H}_{\mathcal{X}_L}\otimes \mathcal{H}_{\mathcal{X}_R}$ is sufficient for $Tr_{\mathcal{H}_{\mathcal{X}}}(e^{-\beta H_{\mathcal{X}}})= Z(\beta)$ to be true, it does not establish that it is necessary. Indeed, in this section we have shown $Tr_{\mathcal{H}_{\mathcal{X}}}(e^{-\beta H_{\mathcal{X}}})= Z(\beta)$  without appealing to factorisation at all. Similarly, \cite{Balasubramanian:2025jeu} showed the fine-grained equality
\beq \label{eq:trfactorisation}
Tr_{\mathcal{H}_{LR}}(e^{-H_{L}\beta_1}e^{-H_{R}\beta_2}) = Z(\beta_1)\times Z(\beta_2) \, 
\eeq
by tracing over a basis of two-sided shell states, also without requiring factorization. The derivation of (\ref{eq:trfactorisation}) in \cite{Balasubramanian:2025jeu} and the derivation of (\ref{Tr=Z}) above  relied instead on the existence of wormhole geometries that separately connect two-boundary or one-boundary states.  By contrast, factorization of the two-boundary Hilbert space (\ref{eq:facprob}) followed from the existence of a {\it different} non-perturbative effect -- ``mixed'' wormholes connecting  two-boundary states to the  one-boundary states \cite{Balasubramanian:2025zey}.   Thus  $Tr_{\mathcal{H}_{\mathcal{X}}}(e^{-\beta H_{\mathcal{X}}})= Z(\beta)$ can follow without factorization of the two-boundary Hilbert space if there is a consistent theory of gravity in which wormholes connecting  only one- or two- boundary states exist but  ``mixed" wormholes do not.  It would be interesting therefore to understand whether or not existence of the first kind of wormhole requires the second for consistency of the theory.

Our analysis in fact suggests the opposite conclusion in theories where the two-boundary Hilbert space does \emph{not} factorise into single-boundary factors, such as matter-free JT gravity and pure 3d gravity. The equality of the Gibbons-Hawking path integral with the single-boundary thermal trace rested on identifying $\mathds{1}_{\mathcal{X}_L}\otimes \mathds{1}_{\mathcal{X}_R}$ with $\mathds{1}_{\mathcal{X}_L \cup \mathcal{X}_R}$, an identification that holds precisely when the two-boundary Hilbert space factorises. When factorisation fails these operators differ, and the Gibbons-Hawking path integral $\overline{\langle\beta|\beta\rangle}=\overline{\langle\beta|\mathds{1}_{\mathcal{X}_L\cup\mathcal{X}_R}|\beta\rangle}$ no longer equals the single-boundary trace $\overline{\langle\beta|\mathds{1}_{\mathcal{X}_L}\otimes\mathds{1}_{\mathcal{X}_R}|\beta\rangle}$. In such theories, then, our results suggest that the Gibbons-Hawking path integral does not admit an interpretation as a thermal trace over the single-boundary Hilbert space at all. This is consistent with the factorisation puzzle of \cite{Harlow:2018tqv}, in which the absence of a factorised Hilbert space obstructs a clean trace interpretation of the gravitational partition function. It would be interesting to understand what, if anything, the Gibbons-Hawking path integral does compute in these cases.

 A corollary of our result is that the exponential of the Bekenstein-Hawking entropy does count the leading order micro-canonical density of states of the gravity theory. We then used the fact that the single-sided shell states span, as shown in \cite{Balasubramanian:2025zey}, to argue that the  basis for the micro-canonical Hilbert space consists solely of black hole states with the same exterior as the Gibbons-Hawking black hole saddle but with a nontrivial interior. Therefore the Euclidean path integral with one boundary and periodic time indeed counts the number of black hole microstates after micro-canonical projection,  a fact that was expected but not previously demonstrated through a direct calculation in gravity. 

 By similar methods we  also arrived at a purely gravitational understanding of why the holographic replica trick  is in fact evaluating the trace required for computing the R\'{e}nyi entropy of gravity states entangled between two universes. As the two-boundary Hilbert space factorises into single-boundary theories, we obtained the entanglement entropy  by tracing out  these factors individually. We showed equality with the holographic replica trick through a general factorisation argument and also by explicitly tracing over the shell basis. We focused our analysis on entanglement between universes with disconnected boundaries. It would be interesting to see if these methods can be extended to tracing out a subregion of a connected boundary component, and to see how our construction is related to the algebraic understanding of gravitational entropy recently discussed in \cite{Colafranceschi:2023moh}.

In this paper we made no assumptions regarding holographic duality and our arguments apply equally in asymptotically AdS and flat spacetime.

\acknowledgments
TY thanks the Peter Davies Scholarship for continued support. VB was supported in part by the DOE through DE-SC0013528 and QuantISED grant DE-SC0020360, and in part by the Eastman Professorship at Balliol College, University of Oxford. VB thanks Simon Ross and Sameer Murthy for discussions about the Gibbons-Hawking approach to black hole entropy.

\bibliographystyle{jhep}
\bibliography{ref}

\end{document}